\documentclass[preprint]{aastex}
\pdfoutput=1

\renewcommand{\vec}[1]{\mathbf{#1}}

\newcommand{\eqref}[1]{(\ref{#1})}
\newcommand{\del}{\vec{\nabla}}
\newcommand{\pder}[2]{\frac{\partial #1}{\partial #2}}
\newcommand{\tder}[2]{\frac{\mathrm{d}#1}{\mathrm{d}#2}}

\shorttitle{Turbulent Mixing of Metals in Galactic Disks}
\shortauthors{Yang \& Krumholz}

\begin{document}

\title{Thermal-Instability-Driven Turbulent Mixing in Galactic Disks:\\
       I.~Effective Mixing of Metals}
       
\author{Chao-Chin Yang and Mark Krumholz}
\affil{Department of Astronomy and Astrophysics, University of California,\\
       1156 High Street, Santa Cruz, CA~95064, U.S.A.}
\email{ccyang@ucolick.org}

\begin{abstract}

Observations show that radial metallicity gradients in disk galaxies are relatively shallow, if not flat, especially at large galactocentric distances and for galaxies in the high-redshift universe. Given that star formation and metal production are centrally concentrated, this requires a mechanism to redistribute metals. However, the nature of this mechanism is poorly understood, let alone quantified. To address this problem, we conduct magnetohydrodynamical simulations of a local shearing sheet of a thin, thermally unstable, gaseous disk driven by a background stellar spiral potential, including metals modeled as passive scalar fields.  Contrary to what a simple $\alpha$ prescription for the gas disk would suggest, we find that turbulence driven by thermal instability is very efficient at mixing metals, regardless of the presence or absence of stellar spiral potentials or magnetic fields.  The timescale for homogenizing randomly distributed metals is comparable to or less than the local orbital time in the disk. This implies that turbulent mixing of metals is a significant process in the history of chemical evolution of disk galaxies.
\end{abstract}


\section{INTRODUCTION}

The spatial distribution of metals in disk galaxies is a crucial clue for understanding how galaxies formed and evolved over cosmic time. The past few decades have produced a wealth of observations of this property, including in our own Milky Way \citep[e.g.,][]{HK10,BR11,LL11,CR12,YCF12}, in nearby galaxies \citep[e.g.,][]{VE92,CC00,PVC04,KC11}, and in the high-redshift universe \citep[e.g.,][]{CM10,JE12,QC12}. In the local universe, a variety of radial metallicity gradients in disk galaxies are seen, but they are generally on the order of $-0.03$~dex~kpc$^{-1}$, with the negative sign indicating decreasing metallicity at larger galactocentric radii. However, surprisingly, these gradients seem to disappear in the outer parts of galactic disks, where there is little star formation; moreover, the metal content is greater than would be expected given the amount of star formation that has taken place at these radii \citep[e.g.,][]{BR09,BKR12,WP11}.  High-redshift galaxies, in comparison, are far less regular. Their metallicity gradients range from negative ones significantly steeper than those found locally, to completely flat or even positive. Any successful theory of galactic evolution must be able to reproduce these observations.

Several processes play an important role in regulating the spatial variation of metals in disk galaxies, and these have been demonstrated by either chemical evolution models or hydrodynamical simulations.  The enrichment of metals in the interstellar medium (ISM) is dominated by star formation and subsequent stellar mass loss, and thus depends on the star formation law \citep[e.g.,][]{PE91}.  Metals are diluted by infalling gas from outside galaxies \citep[e.g.,][]{TL78,cC80,MF89,CMG97,PB00,CMR01}.  Radial inflow of the gas within the disk of a galaxy redistributes metals \citep{MV81,LF85,PT89, GK92,PC00,SM11,BS12}; galaxy interactions are especially effective in inducing large-scale inflow and flattening metallicity gradients (\citealt{RKC10,PMT11}; Torrey et al., in preparation). Turbulence associated with the viscous evolution of gas disks can also redistribute metals \citep{cC89,SY89,SY90,TY95,TM98}. Beyond these gas-dynamical processes, radial migration of stars can alter stellar metallicity distributions independently of processes affecting the gas phase \citep{SB02,RD08a,RD08b,SB09}.

However, the strength and relative importance of these processes remains very poorly understood. Semi-analytic chemical evolution models generally parameterize each process and then tune the parameters in an attempt to provide an acceptable match to observations. However, the large numbers of parameters involved means that even a good fit to the data may not be unique, and the need for fine-tuning means these models have limited predictive power. Moreover, the parameterizations used in the models may not be accurate. For example, turbulent mixing is usually treated by adopting an $\alpha$ prescription for the turbulent transport of angular momentum \citep{SS73}, and assuming that the transport coefficient for metals is the same. Under these assumptions turbulent mixing is unimportant unless $\alpha$ is so large that the viscous diffusion time becomes comparable to the gas depletion time. However, neither the assumption that turbulent transport of metals can be approximated with an $\alpha$ prescription, nor that $\alpha$ for this process is the same as that for the angular momentum, are physically well-motivated. Numerical simulations of metal transport that evolve galaxies over cosmological times are unfortunately little better, because their limited resolution means that they must also adopt parameterized treatments of unresolved processes. For example, most smoothed particle hydrodynamics (SPH) simulations allow no chemical mixing at all between SPH particles \citep{WVC08}, or at best treat mixing approximately using a parameterized subgrid recipe \citep{SWS10}. Eulerian simulations, in contrast, dramatically over-mix when their resolution is low.

The approach we take in this paper is quite different, and complementary to chemical models and large-scale cosmological simulations. We isolate a single process: turbulent mixing within a galactic disk. Our goal is to provide a first-principles calculation of this process, which can in turn provide a physically-motivated, parameter-free prescription that can be used in chemical evolution models or lower resolution simulations. In order to achieve this goal, we simulate turbulent mixing in a portion of a galaxy at very high resolution, including physical processes that are too small-scale to be resolved in cosmological simulations, and we perform a resolution study to ensure that our results are converged. The previous work that most closely matches ours in philosophy and overall approach is that of \citet{MF99} and \citet{FML04}, who used high resolution simulations of isolated portions of galaxies to study mixing of supernova ejecta with the ISM as galactic winds are launched. Here we perform a similar calculation for turbulent mixing within disks.

There exist many sources that can drive turbulence in the ISM (see \citealt{ES04} and \citealt{SE04} and references therein).  In earlier work, \citet{AM02} studied the properties of turbulent mixing driven by supernova explosions.  Here, we instead focus on turbulence driven by thermal instability \citep{gF65, FGH69}. Our motivation is two-fold. First, the flat metallicity gradients seen in outer disks presumably call for some sort of mixing process to operate at large galactic radii, where star formation is limited and thus supernova explosions are extremely rare. Second, even in places where supernovae do occur, thermal instability is also present and will drive turbulence; indeed, thermal instability is essentially inevitable anywhere the ISM is dominated by atomic hydrogen, which is the case for most galaxies over the great majority of cosmic time. Supernovae will only enhance the turbulence compared to what we find, and thus our results should be viewed as a minimum estimate of the turbulent mixing rate.

In the sections that follow, we describe our numerical method and present the results of the simulations. We then quantify the results in a form appropriate for use in chemical evolution models and discuss their implications. Finally, we summarize the results and conclude.

\section{NUMERICAL MODELING}

To study turbulent mixing, we adopt a thin-disk, local-shearing-sheet model similar to \citet{KO02} \citep[see also][]{KO06} and equip it with approximate heating and cooling processes \citep[see also][]{KKO08,KKO10} and metal tracers.  We also investigate the effects of spiral shocks and magnetic fields on turbulent mixing in our models.  In the following sections, we describe our simulation methodology and setup in detail.

\subsection{Governing Equations}

\subsubsection{Magnetohydrodynamics}

Using the local-shearing-sheet approximation \citep{GL65}, we consider a small region distant from the center of a vertically thin disk galaxy.  This region is centered at the potential well of a background stellar spiral arm and co-rotating with the arm at its angular pattern speed $\Omega_p$.  One can define a coordinate system for this region by $x' \equiv R - R_0$ and $y' \equiv R\phi$, where $(R,\phi)$ are the polar coordinates rotating with the spiral arm and $(R,\phi) = (R_0,0)$ is the location of the point of interest along the stellar spiral arm.  The velocity field of the gas flow in the rotating frame is denoted by $\vec{v}$.  Since the differential rotation profile of a disk galaxy $\Omega = \Omega(R)$ is usually a known or given function, we are more interested in the relative velocity of the gas $\vec{u} \equiv \vec{v} - \vec{v}_c$ with respect to the circular velocity $\vec{v}_c \equiv R(\Omega - \Omega_p)\vec{e}_{y'}$ in the rotating frame.  By assuming $x' \ll R_0$, $y' \ll R_0$, $u_{x'} \ll R_0\Omega_0$, and $u_{y'} \ll R_0\Omega_0$, where $\Omega_0 \equiv \Omega(R_0)$, and expanding the magnetohydrodynamical equations to first order in $x'$, $y'$, $u_{x'}$, and $u_{y'}$, the continuity, the momentum, the energy, and the induction equations become
\begin{eqnarray}
  &\pder{\Sigma}{t} + \vec{v}\cdot\del\Sigma + \Sigma\del\cdot\vec{u}
  = 0,\label{E:continuity}\\
  &\pder{\vec{u}}{t} + \vec{v}\cdot\del\vec{u}
  = q_0 \Omega_0 u_{x'} \vec{e}_{y'} - 2\vec{\Omega}_0\times\vec{u}
  - \frac{1}{\Sigma}\del p - \del\left(\Phi_s + \Phi_g\right)
  + \frac{1}{\Sigma}\vec{J}\times\left(\vec{B}_0 + \vec{B}\right),
    \label{E:momentum}\\
  &\pder{e}{t} + \vec{v}\cdot\del e + e\del\cdot\vec{u}
  = -p\del\cdot\vec{u} + \mathcal{H},\label{E:energy}\\
  &\pder{\vec{A}}{t} + \vec{v}_c\cdot\del\vec{A}
  = q_1\Omega_0 A_{y'}\vec{e}_{x'}
  + \vec{u}\times\left(\vec{B}_0 + \vec{B}\right),\label{E:induction}
\end{eqnarray}
respectively.  The primitive variables for which we solve the above equations are the gas surface density $\Sigma$, the gas relative velocity $\vec{u}$ as defined above, the thermal energy density of the gas $e$ (i.e., internal energy per unit surface area), and the magnetic vector potential $\vec{A}$.  The magnetic field $\vec{B}$ is then calculated by $\vec{B} = \del\times\vec{A}$.  The remaining quantities in the above equations are the dimensionless shear parameters
\begin{eqnarray}
  q_0 &\equiv& -\left.\frac{R}{\Omega}\tder{\Omega}{R}\right|_{R = R_0},\label{E:q_0}\\
  q_1 &\equiv& -\left.\frac{1}{\Omega}
                \tder{R\left(\Omega - \Omega_p\right)}{R}\right|_{R = R_0}
            = q_0 - 1 + \frac{\Omega_p}{\Omega_0},\label{E:q_1}
\end{eqnarray}
the gas pressure $p$, the gravitational potentials due to the stellar spiral arm and the gas itself, $\Phi_s$ and $\Phi_g$, respectively, the electric current density $\vec{J} = (\del\times\vec{B}) / \mu_0$, where $\mu_0$ is the permeability, and the net heating rate per unit surface area $\mathcal{H}$.  We impose a constant external azimuthal magnetic field $\vec{B}_0 = B_0\vec{e}_{y'}$ when we consider a magnetized disk.\footnote{The induction equation~\eqref{E:induction} requires that $\vec{B}_0 \parallel \vec{v}_c$; otherwise, an additional term $\vec{v}_c \times \vec{B}_0$ should be included.}

Equations~\eqref{E:continuity}--\eqref{E:induction} are written in vectorial forms and thus are readily transformed into different coordinate systems.  One particular choice is to rotate the $(x', y')$ system counterclockwise by the pitch angle $i$ into the $(x, y)$ system such that the new $x$-axis and $y$-axis are perpendicular and parallel to the stellar spiral arm at the origin, respectively \citep[see Figure~1 of][]{KO02}.  We employ the tightly-wound approximation so that $\sin i \approx i \ll 1$ can be deemed a small quantity.  In these considerations, Equations~\eqref{E:continuity} and~\eqref{E:energy} remain unchanged while Equations~\eqref{E:momentum} and~\eqref{E:induction} become, by also retaining $\sin i$ to only first order,
\begin{eqnarray}
  &\pder{\vec{u}}{t} + \vec{v}\cdot\del\vec{u}
  = q_0 \Omega_0 u_x \vec{e}_y - 2\vec{\Omega}_0\times\vec{u}
  - \frac{1}{\Sigma}\del p - \del\left(\Phi_s + \Phi_g\right)
  + \frac{1}{\Sigma}\vec{J}\times\left(\vec{B}_0 + \vec{B}\right),\label{E:momentum_1}\\
  &\pder{\vec{A}}{t} + \vec{v}_c\cdot\del\vec{A}
  = q_1\Omega_0\left[\left(A_x\sin i + A_y\right)\vec{e}_x
                   - A_y\vec{e}_y\sin i\right]
  + \vec{u}\times\left(\vec{B}_0 + \vec{B}\right),
  \label{E:induction_1}
\end{eqnarray}
respectively.  The circular velocity and the external magnetic field in this tilted frame can be approximated by
\begin{eqnarray}
  &\vec{v}_c \approx v_0\vec{e}_x\sin i + \left(v_0 - q_1\Omega_0 x\right)\vec{e}_y,\label{E:cirvel}\\
  &\vec{B}_0 \approx B_0\vec{e}_x\sin i + B_0\vec{e}_y,
\end{eqnarray}
respectively, where $v_0 \equiv R_0\left(\Omega_0 - \Omega_p\right)$.

\subsubsection{Forcing Driven by the Stellar Spiral Arm} \label{SS:forcing}

The advantage of aligning our coordinate system with the background stellar spiral arm is that the gravitational potential of the arm in this system can be approximated as periodic in $x$ while weakly varying in $y$, to first order \citep{wR69,SMR73,KO02}:
\begin{equation} \label{E:phis}
  \Phi_s(x) \approx \Phi_0\cos\frac{2\pi x}{L},
\end{equation}
where $\Phi_0 < 0$ is a constant,
\begin{equation}
  L = \frac{2\pi R_0\sin i}{m} \label{E:spacing}
\end{equation}
is the radial spacing between adjacent spiral arms, and $m$ is the multiplicity of the arms.  The strength of the spiral forcing is measured in terms of the local centrifugal acceleration:
\begin{equation}
  F \equiv \frac{m}{\sin i}\left(\frac{\left|\Phi_0\right|}{R_0^2\Omega_0^2}\right).\label{E:force}
\end{equation}
Note that according to Equation~\eqref{E:spacing}, the local-shearing-sheet approximation $L_x \ll R_0$ requires that $\sin i / m \ll 1$, which is automatically satisfied by the tightly-wound approximation $\sin i \ll 1$.

\subsubsection{Self-gravity of the Gas}

In this work, we are only interested in large-scale mixing of metals and ignore self-gravity of the gas.  It will be required, though, in a subsequent paper where we investigate the metal abundances in precursors of molecular clouds.  Therefore, we list the equation governing self-gravity of the gas in this section for completeness.

The quantity $\Phi_g$ in Equations~\eqref{E:momentum} and~\eqref{E:momentum_1} represents the gravitational potential of the gas.  The potential is calculated by solving the Poisson equation for a razor-thin disk
\begin{equation}
  \nabla^2\Phi_g = 4\pi G \Sigma \delta(z),\label{E:poisson}
\end{equation}
where $G$ is the gravitational constant and $\delta(z)$ is the Dirac delta function.  We discuss the corresponding numerical method for its solution in Section~\ref{SS:pc}.

\subsubsection{Thermodynamics}

To understand the effects of thermal instability on the mixing of metals, we compare thermally stable disks with thermally unstable ones.  For the former, we use the isothermal equation of state $p = c_s^2 \Sigma$, where $c_s$ is the isothermal speed of sound, and in this case, the energy equation~\eqref{E:energy} is not required and we do not solve it.  For the latter, we adopt the adiabatic equation of state $p = (\gamma - 1)e$, where $\gamma$ is the two-dimensional adiabatic index, and include the heating and cooling of the gas such that the disk is thermally unstable.

For a prescription of the heating and cooling rates, we start from the approximate functions suggested by \citet{KI02}.\footnote{The cooling function published in \citet{KI02} contains two typographical errors and has been corrected by \citet{NIK06}.}  The net rate of heat loss per unit volume is $\rho\mathcal{L} = n^2\Lambda - n\Gamma$, where $n$ is the number density of gas particles and
\begin{eqnarray}
  \Gamma &=& 2.0\times10^{-26}~\textrm{erg s}^{-1},\\
  \frac{\Lambda(T)}{\Gamma}
  &=& 10^7\,\exp\left(-\frac{1.184\times10^5}{T+1000}\right)
  + 1.4\times10^{-2}\,\sqrt{T}\exp\left(-\frac{92}{T}\right)~\textrm{cm}^3,
\end{eqnarray}
in which $T$ is the temperature in Kelvins.  To obtain the heating rate per unit surface area $\mathcal{H}$, we need to integrate $\rho\mathcal{L}$ in the vertical direction, and the vertical structure of the gas is required.  For simplicity, we assume that the gas is vertically isothermal and the number density is approximated by $n(z) \simeq n_0\exp\left(-z^2 / 2H^2\right)$, where $n_0$ is the number density in the mid-plane and $H$ is the vertical scale height.  We further assume that the vertical motion of the gas is dominated by non-thermal processes, at least when the gas temperature is low, such that $H$ is constant, instead of depending on $T$.  Therefore,
\begin{equation} \label{E:heating}
  \mathcal{H} = -\int \rho\mathcal{L}\,\mathrm{d}z
              = \Gamma\left(\frac{\Sigma}{\mu m_u}\right)
                \left[1 - \frac{1}{2\sqrt{\pi}H}
                          \left(\frac{\Sigma}{\mu m_u}\right)
                          \frac{\Lambda(T)}{\Gamma}\right],
\end{equation}
where we have used $\int n\,\mathrm{d}z = \Sigma / \mu m_u$, and $\mu$ and $m_u$ are the mean molecular weight and the atomic mass, respectively.  To complete the system, we use the ideal-gas law $p = \Sigma k_B T / \mu m_u$ for the temperature $T$, where $k_B$ is the Boltzmann constant.

\subsubsection{Metal Tracers}

Finally, we assume the heavy metals in the disk follow the velocity field of the gas component and model them as tracer fluids.  Therefore, the surface density of any given metal $\Sigma_X$ satisfies
\begin{equation}
  \pder{\Sigma_X}{t} + \vec{v}\cdot\del\Sigma_X + \Sigma_X\del\cdot\vec{u}
  = 0.\label{E:tracer}\\
\end{equation}
The concentration of the metal $c$ with respect to the total local mass can then be derived by $c = \Sigma_X / \Sigma$.

\subsection{Initial and Boundary Conditions} \label{SS:ibc}

The computational domain we consider is a square sheet of size $L$, the radial spacing between adjacent spiral arms defined in Section~\ref{SS:forcing}.  In the following subsections, we discuss the initial and boundary conditions we adopt for the gas and the metals.

\subsubsection{The Gas and the Equilibrium State}

We initially set the gas to be uniform ($\Sigma = \Sigma_0$), isothermal ($T = T_0$), and moving along with the galactic circular motion ($\vec{u} = 0$) such that it is at an equilibrium state when the spiral forcing is not present ($\Phi_0 = F = 0$).  Our adopted initial density $\Sigma_0$ can be more physically motivated by considering the corresponding Toomre $Q$ parameter for the gas $Q_0 = \kappa c_{s,0} / \pi G \Sigma_0$, where $\kappa$ is the epicycle frequency of the disk at $R = R_0$ and $c_{s,0}$ is the initial speed of sound.  The epicycle frequency $\kappa$ as well as the shear parameters $q_0$ and $q_1$ defined in Equations~\eqref{E:q_0} and~\eqref{E:q_1} are determined by the rotation profile $\Omega(R)$.  As in \citet{KO02}, we assume a flat rotation curve ($R\Omega = \textrm{constant}$) near $R = R_0$ and a pattern speed of $\Omega_p = \Omega_0 / 2$ and thus $\kappa = \sqrt{2}\Omega_0$, $q_0 = 1$, and $q_1 = 1/2$.  In physical units, the initial surface density is then
\begin{equation}
  \Sigma_0 = \left(19~M_\sun~\textrm{pc}^{-2}\right) Q_0^{-1}
                       \left(\frac{\Omega_0}{26~\textrm{km}~\textrm{s}^{-1}~\textrm{kpc}^{-1}}\right)
                       \left(\frac{c_{s,0}}{7.0~\textrm{km}~\textrm{s}^{-1}}\right).
\end{equation}
On top of the equilibrium state, we perturb the velocity field by a white noise of magnitude $10^{-3}c_{s,0}$ to seed the instabilities, if any, of the system.

\begin{figure}[!tbp]
\begin{center}
\plotone{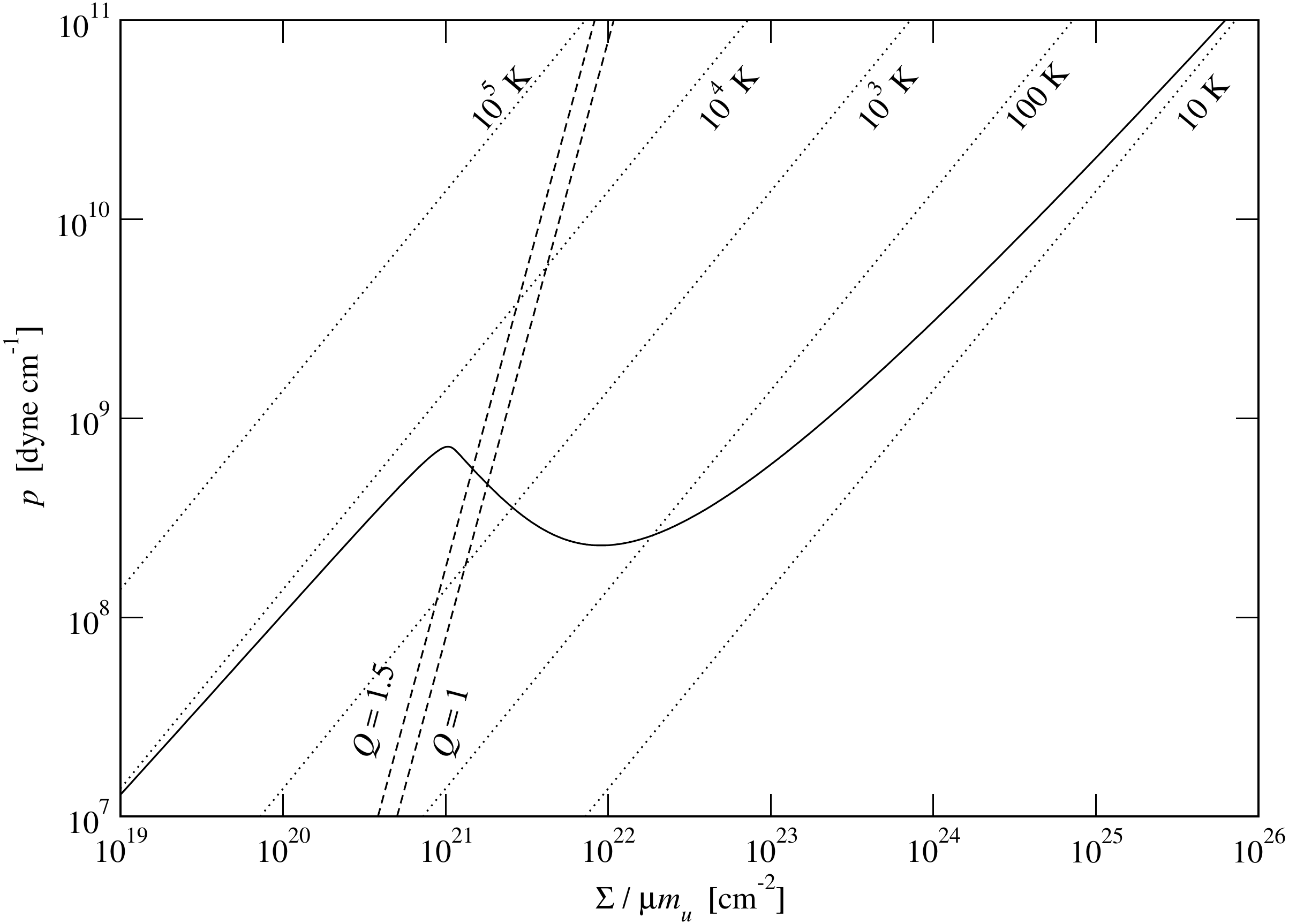}
\caption{Pressure-density diagram for non-isothermal gas in our models.  The \emph{solid} line indicates the states at thermal equilibrium, the \emph{dotted} lines show the isotherms, and the \emph{dashed} lines denote constant Toomre stability parameter for the gas.}
\label{F:theq}
\end{center}
\end{figure}

When we consider the isothermal equation of state, our initial isothermality of the gas is automatically guaranteed and preserved.  In this case, only the constant speed of sound $c_s = c_{s,0}$ needs to specified.  On the other hand, when we consider a non-isothermal disk with heating and cooling processes, an initial thermal equilibrium of the gas is also required.  This is equivalent to setting $\mathcal{H} = 0$ at $\Sigma = \Sigma_0$ and $T = T_0$, i.e., zero net heating rate at the initial state.  Given a surface density $\Sigma$, Equation~\eqref{E:heating} can be used to solve for the corresponding temperature $T$ such that $\mathcal{H} = 0$.  With the values of the physical parameters considered in this work (see Section~\ref{SSS:param} and Table~\ref{T:param}), Figure~\ref{F:theq} plots the curve for the states at thermal equilibrium in a pressure-density diagram; notice the region in $\Sigma / \mu m_u \sim 10^{21}$--$10^{22}$~cm$^{-2}$ where the slope of the curve is inverted, the classical condition for thermal instability to occur \citep{gF65}.  Therefore, a given value of initial Toomre stability parameter for the gas $Q_0$ uniquely specifies the initial state of the gas $(\Sigma_0, T_0)$.

For magnetized disks, we impose initial uniform azimuthal fields through the gas by setting $B_0 \neq 0$ and $\vec{A} = 0$ throughout.  The strength of these imposed fields can be gauged by the corresponding plasma beta parameter, $\beta_0 \equiv 2\mu_0 p_0 / B_0^2$, which is the ratio of the initial thermal pressure $p_0$ to the initial magnetic pressure.

Since our system is driven by a forcing periodic in the $x$ direction with a wavelength of the inter-arm spacing $L$ (Equation~\eqref{E:phis}), it is expected that the response of the system also be periodic in $x$ with the same wavelength.  The system is also sheared in the $y$ direction, though, which is manifested by the term $-q_1 \Omega_0 x \vec{e}_y$ in Equation~\eqref{E:cirvel}.  So strictly speaking, the boundary conditions in the $x$ direction for the system should be sheared periodic, which can be expressed mathematically in the form $f(x + L, y) = f(x, y + q_1 \Omega_0 L t)$, where $f(x,y)$ is any dynamical field in question \citep{HGB95,BN95,KO02}, except the metal tracer fields (see below).  As for the $y$ direction, we adopt normal periodic boundary conditions $f(x, y + L) = f(x,y)$ for convenience.

\subsubsection{The Metal Tracers} \label{SSS:metals}

Given that we model the metals as passive scalar fields, they effectively act like dye in a flow.  We can in fact inject metals anywhere, anytime, and at any rate, to study their diffusion process.  Since most star formation occurs along spiral arms, constantly producing metals that drift downstream towards the next spiral arm, we are most interested in the mixing of metals within one passage between adjacent arms.  This motivates us to employ inflow boundary conditions at the left ($x = -L/2$) and outflow boundary conditions at the right ($x = +L/2$) for the metal tracer field $\Sigma_X$.  The boundary conditions in the $y$ direction remains periodic.

The spatial distribution of the newly produced metals from the previous generation of star formation may be arbitrary.  To be as general as possible, we constantly inject a sinusoidal distribution of unit amplitude from the left: $\Sigma_X(x=-L/2, y) = \sin(2\pi y / \lambda_\mathrm{inj})$, where $\lambda_\mathrm{inj}$ is the wavelength of the distribution.  Once the wavelength dependence of the mixing process is deciphered, an arbitrary distribution of metals can be analyzed by Fourier decomposition.  In all of our models, we simultaneously evolve four species of metal tracers with $\lambda_\mathrm{inj} = L, L/2, L/4$, and $L/8$, for which we denote $X$ by 1, 2, 3, and 4, respectively.

At the equilibrium state when there exists no driving force in the system, the gas flows azimuthally at circular velocity, and so do the metals.  Given that the azimuthal direction in our computational domain is tilted from the $y$-axis by the pitch angle $i$, we set the initial condition for the metal tracers as
\begin{equation}
  \Sigma_X(x,y) = \sin\left[\frac{2\pi}{\lambda_\mathrm{inj}}\left(y - \frac{x + L/2}{\tan i}\right)\right].
\end{equation}

\subsubsection{Physical Parameters and the Models} \label{SSS:param}

The values of the physical parameters we adopt and keep constant across different models are listed in Table~\ref{T:param}.  Most of these values match those used by \citet{KO02} for comparison purposes.  The additional two parameters, mean molecular weight $\mu$ and vertical scale height $H$, come into the system only via the thermodynamics.  For simplicity, we set $\mu = 1$ throughout.\footnote{A more realistic value should be $\mu \simeq 1.3$.  But this difference does not significantly change our results.}  The vertical scale height is estimated by noting that $H = c_{s,z} / \nu$ in a vertically isothermal gas with effective vertical speed of sound $c_{s,z}$ and vertical frequency $\nu$; in the Solar neighborhood, $\nu \simeq 2\Omega$ \citep{BT08}.  By assuming $c_{s,z} \simeq 7.0$~km~s$^{-1}$, we calculate $H \simeq 95$~pc.

We conduct separate simulations with each combination of three physical effects --- spiral forcing, thermal instability, and magnetic fields --- to study their influence on the gas dynamics and more importantly the mixing of metals.  The eight resulting models and their numerical resolutions are listed in Table~\ref{T:model}. We show in Appendix \ref{S:res} that this resolution is sufficient to achieve numerically-converged results.  If a model includes spiral forcing, we gradually increase its strength for a duration of 10$P$ to $F = 3$\%, after which the strength remains constant.  When considering the isothermal equation of state, we use a speed of sound of $c_{s,0} = 7.0$~km~s$^{-1}$ and thus an initial gas surface density of $\Sigma_0 = 13~M_\sun$~pc$^{-2}$.  When considering thermally unstable gas, we adopt a two-dimensional adiabatic index of $\gamma = 1.8$, which is taken to be the limiting value of a strongly self-gravitating disk of monatomic gas \citep{cG01}.  The initial thermal equilibrium in this case requires that $\Sigma_0 = 12~M_\sun$~pc$^{-2}$ and $c_{s,0} = 6.4$~km~s$^{-1}$.  For magnetized disks, we set the initial plasma beta to be $\beta_0 = 2$.

\subsection{The Pencil Code} \label{SS:pc}

We use the Pencil Code\footnote{The Pencil Code is publicly available at \texttt{http://code.google.com/p/pencil-code/}.} to solve our system of equations discussed above.  It is a cache-efficient, parallelized code optimal for simulating compressible turbulent flows.  It solves the MHD equations, among others, by sixth-order finite differences in space and third-order Runge-Kutta steps in time, attaining high fidelity at high spectral frequencies \citep{aB03}.  Although the scheme is not written in conservative form, conserved quantities are monitored to assess the quality of the solution.

Several diffusive operations are employed in order to stabilize the scheme.  We use hyper-diffusion in all the four dynamical Equations~\eqref{E:continuity}, \eqref{E:energy}, \eqref{E:momentum_1}, and~\eqref{E:induction_1} to damp noise near the Nyquist frequency while preserving power on most of the larger scales \citep{HB04,JK05}.  Shocks are controlled with artificial diffusion of von Neumann type \citep{HBM04,LJ08}.  For both types of operations, we fix the mesh Reynolds number to maintain roughly the same strength of diffusion at the grid scale (see Appendix~\ref{S:fixRe}).  Finally, all the advection terms of the form $\left(\vec{v}_0 + \vec{u}\right)\cdot\del\mathcal{Q}$, where $\vec{v}_0 = v_0\vec{e}_x\sin i + v_0\vec{e}_y$ (see Equation~\eqref{E:cirvel}) and $\mathcal{Q}$ is any state variable, are treated by fifth-order upwinding to avoid spurious oscillations near stagnation points \citep{DSB06}.

Since a local shearing sheet is considered, we need to handle the sheared advection, the boundary conditions, and the Poisson equation with care.  The sheared advection terms of the form $-q_1\Omega_0 x\partial_y\mathcal{Q}$ are directly integrated by Fourier interpolations \citep{JYK09} in order to relieve the time step constraint from shearing velocity \citep{cG01} and eliminate the artificial radial dependence of numerical diffusion \citep{JGG08}.  The sheared periodic boundary conditions discussed in Section~\ref{SS:ibc} are similarly implemented with Fourier interpolations.  The Poisson Equation~\eqref{E:poisson} is solved by fast Fourier transforms in sheared Fourier space in which the fields are strictly periodic \citep{JO07}.

We have implemented the approximate net heating function (Equation~\eqref{E:heating}) in the Pencil Code.  Since the thermal and the dynamical timescales can be quite different, we operator split this term in the energy Equation~\eqref{E:energy}.  Because this heating and cooling process only depends on local properties, the resulting differential equation is ordinary and can be integrated independently at each cell.  For these integrations, we adopt the fifth-order embedded Runge-Kutta method with adaptive time steps \citep{PF92}.  Since most computational cells are near thermal equilibrium, they require only one or two iterations to match the hydrodynamical time step. Integration of the remaining few cells that have shorter thermal times has negligible computational cost.

With our highest resolution of $\sim$1.5~pc per cell, we are still not able to resolve the thermally stable cold phase of the gas (see Figure~\ref{F:theq}).  Therefore, the densest cells tend to overcool and lose pressure support to their surroundings.  To ensure the Jeans length is properly resolved and avoid artificial fragmentation \citep{TK97} later when we include self-gravity of the gas, we impose a floor to thermal energy density $e$ in accord with the local surface density $\Sigma$ such that the condition of at least four cells per Jeans length is satisfied: $e \ge 4G\Sigma^2 h / \gamma(\gamma - 1)$, where $h$ is the cell size.  This is in effect a modification to the cooling function at low temperatures.  Given that our cell size is marginally close to resolving the stable cold phase and our main purpose is to demonstrate if thermal instability can drive effective chemical mixing, we omit any further consideration of sub-scale physics and leave this compromise as a caveat.

\section{STATISTICALLY STEADY STATE}

All of our models attain statistically steady state within a few local orbital periods.  In this section, we report the state of our simulations at this stage.

\subsection{Gas Dynamics} \label{SS:gas}

\begin{figure}[!tbp]
\begin{center}
\plotone{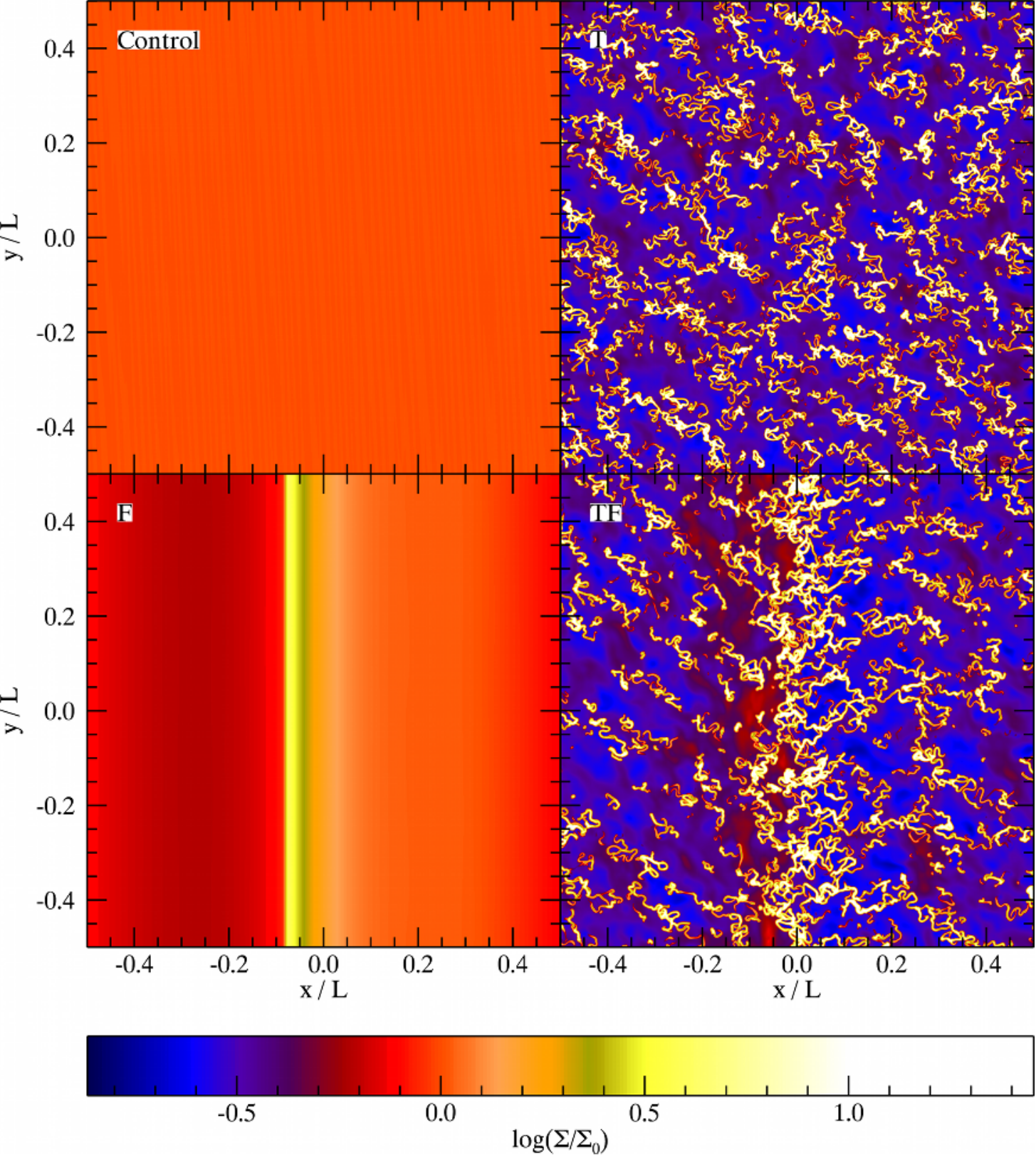}
\caption{Snapshot of the density field for each non-magnetic model at $t = 15P$.  Isothermal disks are in the left column, while thermally unstable disks are in the right column.  The top row has no spiral forcing, while the bottom row does.  The color scales are set the same for all the panels.}
\label{F:rho_nomag}
\end{center}
\end{figure}

\begin{figure}[!tbp]
\begin{center}
\epsscale{.95}
\plotone{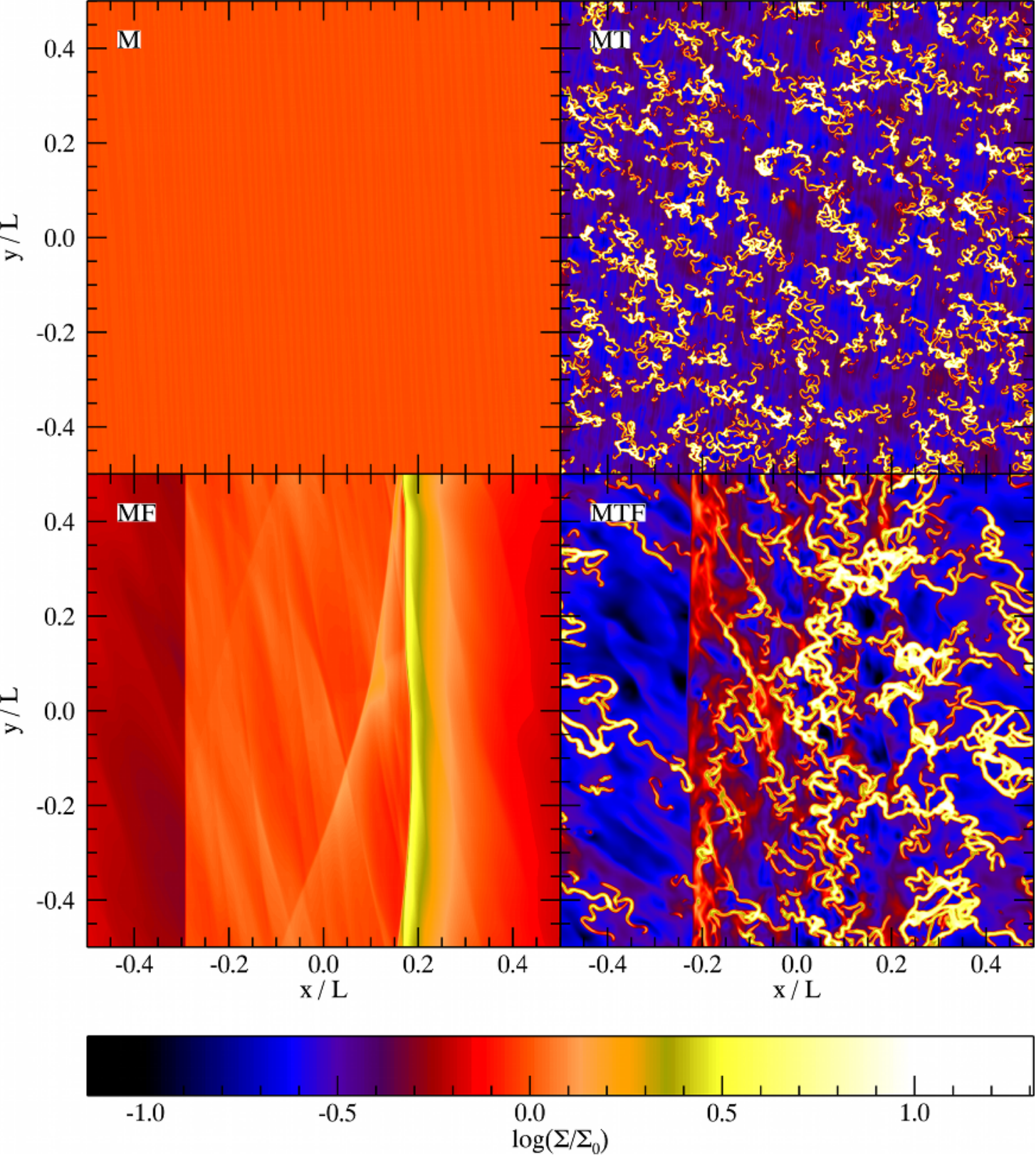}
\caption{Snapshot of the density field for each magnetic model at $t = 15P$.  The arrangement is the same as in Figure~\ref{F:rho_nomag}.}
\label{F:rho_mag}
\end{center}
\end{figure}

\begin{figure}[!tbp]
\begin{center}
\plotone{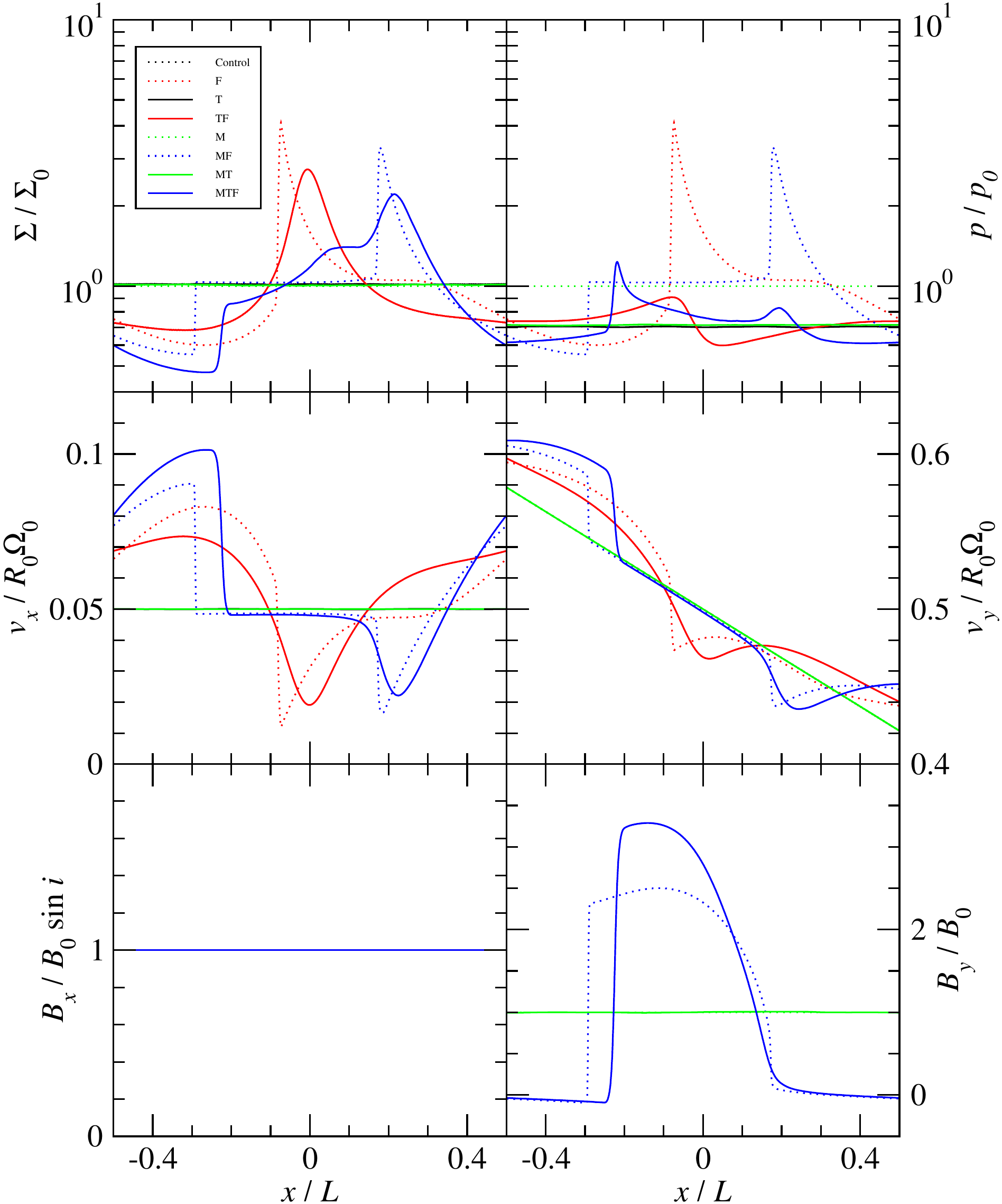}
\caption{The $y$-averages of the density, velocity, and magnetic fields as a function of $x$.  These profiles are also time averaged from $t = 10P$ to $t = 15P$.}
\label{F:yaver}
\end{center}
\end{figure}

Figures~\ref{F:rho_nomag} and~\ref{F:rho_mag} show the snapshots of the density field for the non-magnetic and magnetic models, respectively.  For models~Control and~M, since there exists no driving force in the system (i.e., neither spiral forcing nor thermal instability), no interesting feature occurs and these models serve as control simulations; the initial perturbation propagates as sonic waves and remains small in amplitude.  For model~F, isothermal disk with spiral forcing, a spiral shock with little azimuthal variation forms and locates slightly upstream of the potential well of the forcing.  As can be more clearly seen in the $y$-averages of the gas properties plotted in Figure~\ref{F:yaver}, this resembles the classical solution of \citet{wR69}.

When the disk is thermally unstable, the gas spontaneously breaks into two phases, a warm phase of diffuse gas with high volume filling factor and a cold phase of dense gas with filamentary structures, as evident in models~T and~TF shown in Figure~\ref{F:rho_nomag}.  Model~T is simply the standard two-phase model regulated by thermal instability \citep{FGH69}.  The two-phase medium is turbulent but statistically steady, in which the two phases are in approximate pressure equilibrium.  As demonstrated by model~TF, although the spiral forcing tends to concentrate material into the potential well, the spiral shock seen in isothermal disks is suppressed by the turbulent medium (Figure~\ref{F:yaver}).

The existence of magnetic fields creates an interesting structure in the gas.  For model~MF, magnetized isothermal disk with spiral forcing, two discontinuities parallel to the spiral arm occur as shown in Figures~\ref{F:rho_mag} and~\ref{F:yaver}.  The gas remains at roughly the initial state in between the discontinuities, where the magnetic field lines are compressed and magnetic pressure is increased.  While the left discontinuity is quite stable, the right discontinuity becomes wobbly after the spiral forcing reaches its maximum strength.  Waves are produced in the process and they propagate throughout the disk.  This behavior, however, does not continue to develop in magnitude and drive the gas into a turbulent state.

Finally, the magnetized, two-phase, turbulent medium is rather similar to its non-magnetized counterpart, as can be seen by comparing model~MT shown in Figure~\ref{F:rho_mag} with model~T shown in Figure~\ref{F:rho_nomag}.  As in the non-magnetized case, the presence of thermal instability significantly weakens the spiral shock, as demonstrated by model~MTF shown in Figures~\ref{F:rho_mag} and~\ref{F:yaver}.

\subsection{Metal Tracers}

\begin{figure}[!tbp]
\begin{center}
\plotone{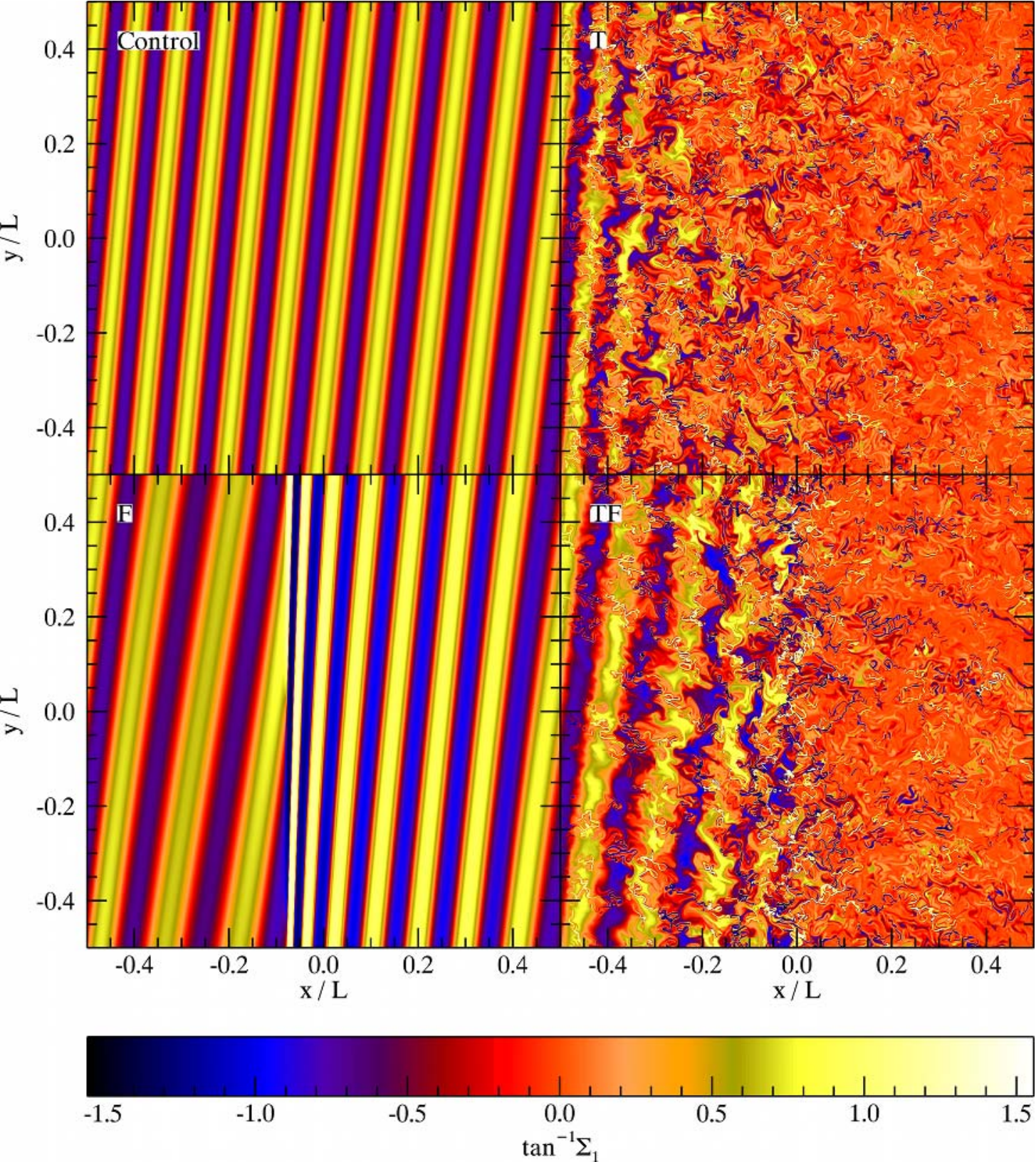}
\caption{Snapshot of the metal tracer field with an injection wavelength of $\lambda_\mathrm{inj} = L$ for each non-magnetic model at $t = 15P$.  The arrangement is the same as in Figure~\ref{F:rho_nomag}.}
\label{F:cc1_nomag}
\end{center}
\end{figure}

\begin{figure}[!tbp]
\begin{center}
\plotone{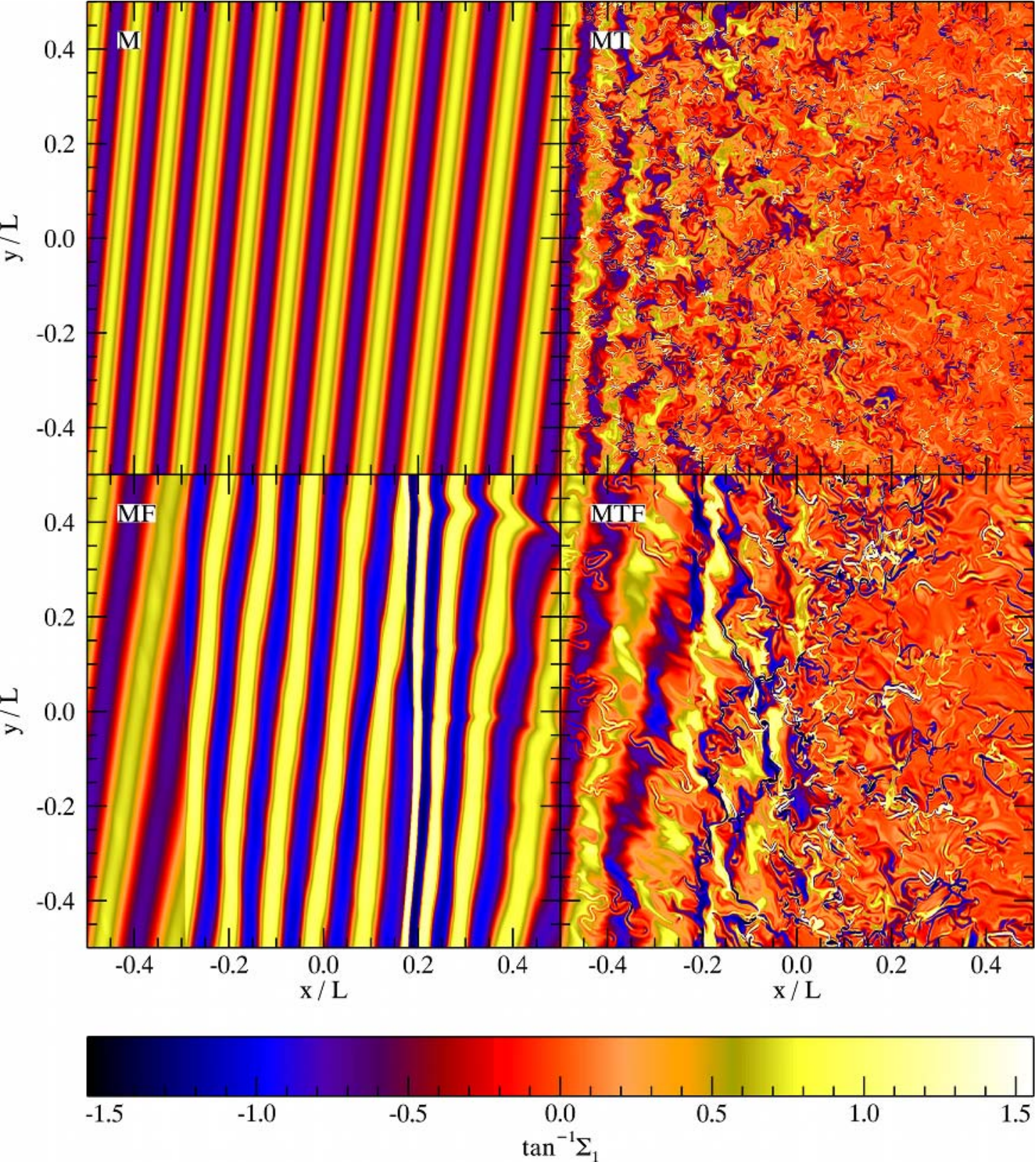}
\caption{Snapshot of the metal tracer field with an injection wavelength of $\lambda_\mathrm{inj} = L$ for each magnetic model at $t = 15P$.  The arrangement is the same as in Figure~\ref{F:rho_mag}.}
\label{F:cc1_mag}
\end{center}
\end{figure}

\begin{figure}[!tbp]
\begin{center}
\plotone{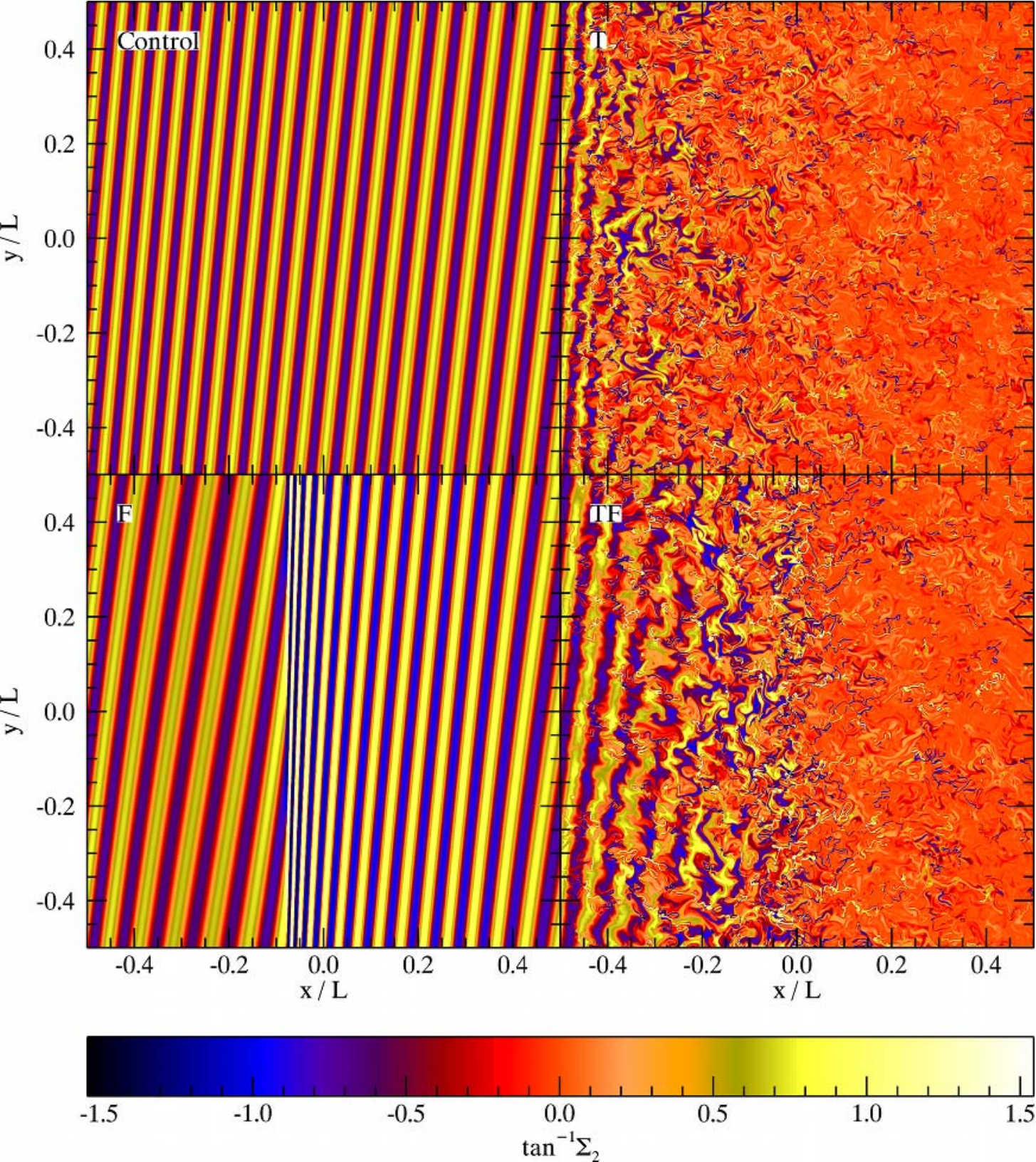}
\caption{Snapshot of the metal tracer field with an injection wavelength of $\lambda_\mathrm{inj} = L / 2$ for each non-magnetic model at $t = 15P$.  The arrangement is the same as in Figure~\ref{F:rho_nomag}.}
\label{F:cc2_nomag}
\end{center}
\end{figure}

\begin{figure}[!tbp]
\begin{center}
\plotone{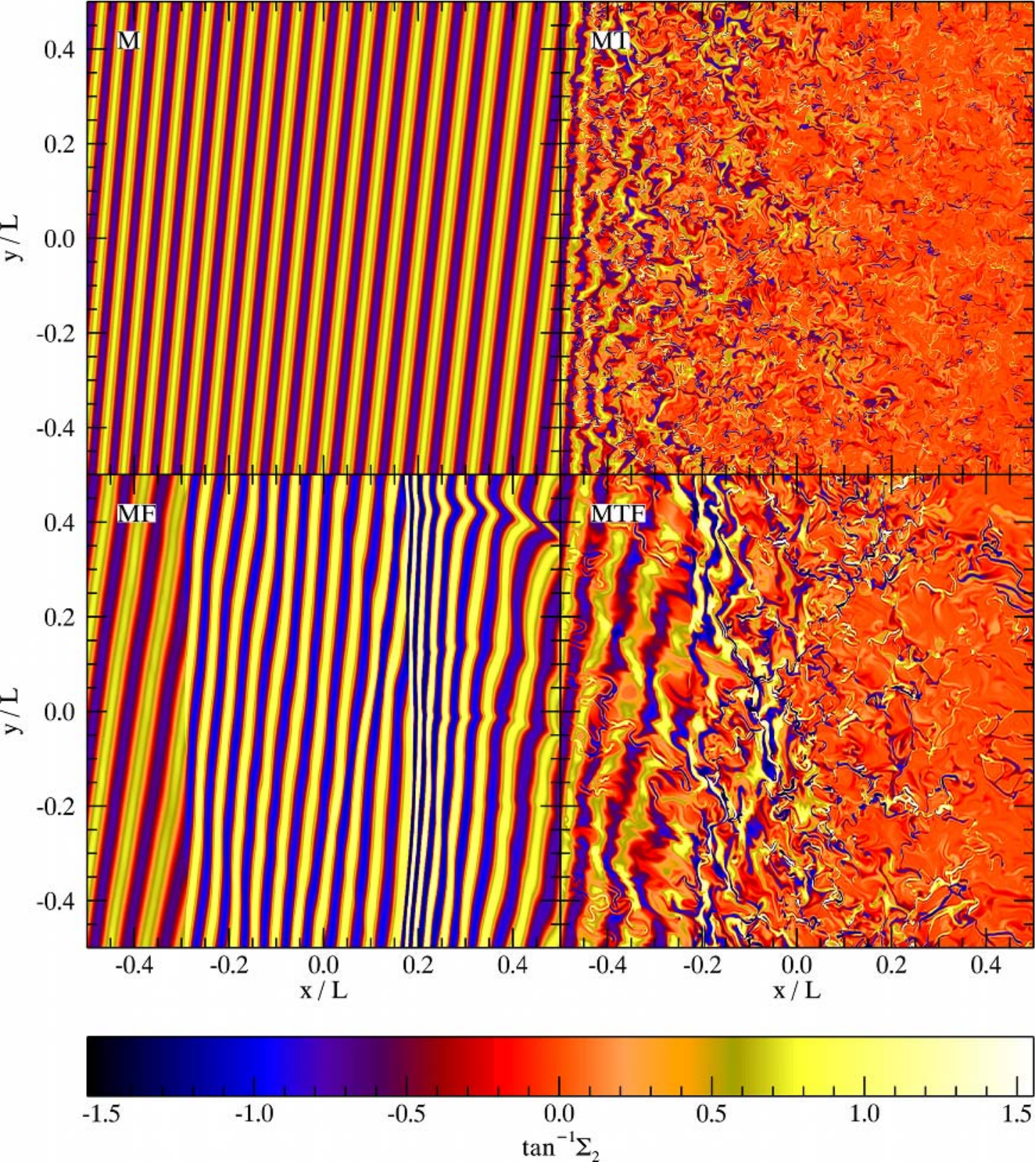}
\caption{Snapshot of the metal tracer field with an injection wavelength of $\lambda_\mathrm{inj} = L / 2$ for each magnetic model at $t = 15P$.  The arrangement is the same as in Figure~\ref{F:rho_mag}.}
\label{F:cc2_mag}
\end{center}
\end{figure}

With a statistically steady state of the gas established for each model listed in Table~\ref{T:model}, we turn to observe how metals would be transported in each flow.  Figures~\ref{F:cc1_nomag}--\ref{F:cc2_mag} show snapshots of the metal tracer fields with an injection wavelength of $\lambda_\mathrm{inj} = L$ or $L/2$ for each model at time $t = 15P$.  Since models~Control and~M remain at their initial equilibrium states, as described in Section~\ref{SS:gas}, we expect the metal tracers do the same.  In this case, metals should move in the azimuthal direction (which is tilted to the right at an angle $i$ with respect to the $y$-axis) without any noticeable diffusion,\footnote{We note that molecular diffusion is too small to drive metal diffusion on galactic scale, and this process is obviously ignored in our models.} and our simulations have passed this benchmark as demonstrated in the snapshots.

When the spiral forcing is present in our non-magnetized isothermal disk (model~F), the tracer field is deformed in the $x$ direction, which is perpendicular to the spiral arm.  In accord with the velocity field shown in Figure~\ref{F:yaver}, the metal distribution is first rarified and its wavelength is increased, as the gas approaches the spiral shock.  As the gas passes through the shock, the metal tracer amplitude is increased to $\sim$5.5 and the wavelength is reduced.  Finally the metal field is rarified again to regain the original distribution towards the right boundary.  A similar process occurs in our magnetized isothermal disk (model~MF), except that in this case there exist two discontinuities and propagating waves generated by the wobbly shock on the right.  The waves, however, do not have enough strength to stir the metals, and the metals  again tend to retain their original distribution after crossing the right shock.  Therefore, the spiral forcing in our isothermal disks, either non-magnetized or magnetized, does not have a noticeable net effect on metal distribution after one passage of the spiral arm.

\begin{figure}[!tbp]
\begin{center}
\plotone{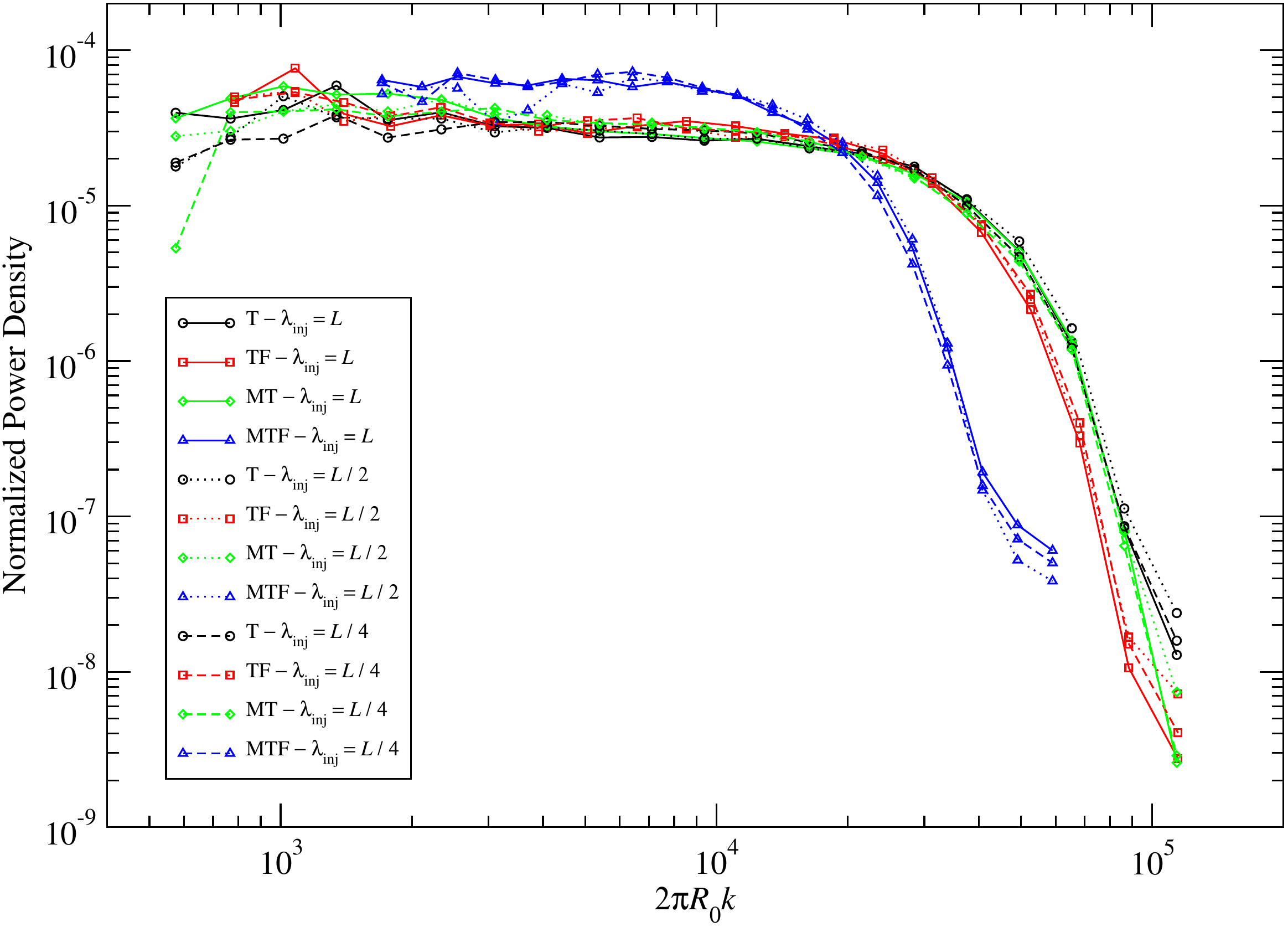}
\caption{Power spectra of the metal tracer fields with different injection wavelengths $\lambda_\mathrm{inj}$ in the downstream region of our thermally unstable disks.  The downstream region is selected as the domain $x > 0$ if no spiral forcing is present, or the aftershock region determined from Figure~\ref{F:yaver} if spiral forcing is present.  The spectra are smoothed in 20 radial logarithmic bins and normalized by $\int\left|\tilde{\Sigma}_X(k)\right|^2\mathrm{d}k = 1$, where $\tilde{\Sigma}_X(k)$ is the Fourier amplitude of the tracer field $\Sigma_X$ at wavenumber $k$.  The values of $k$ at which the power spectra begin to decrease are the Nyquist frequencies in the simulations; the turndown is at lower $k$ in model MTF due to the lower resolution of this model.
\label{F:cc_power}
}
\end{center}
\end{figure}

A completely different scenario for transporting metals occurs in thermally unstable disks.  As evident in model~T, the turbulence driven by thermal instability significantly churns up the metals, and within only a few wavelengths in distance, the original sinusoidal distribution cannot be discerned anymore.
To quantify this process, we compute the power spectra of the metal tracer fields at our final times,
\begin{equation}
  P_X(k) = |\tilde{\Sigma}_X(k)|^2,
\end{equation}
where $\tilde{\Sigma}_X$ is the Fourier transform of a given metal tracer field. We compute the Fourier transform and thus the power spectrum only for gas in the downstream region, defined as the region $x>0$ for the runs without spiral arm forcing, and as the region beyond the spiral shock for runs with forcing. For models Control and M, the power spectrum is simply a $\delta$ function at the injection wavelength, while for model F it is a $\delta$ function at a wavelength smaller than the injection scale (due to compression of the wavelength in the spiral shock). Model MF is not quite a $\delta$ function, but is nearly one. In contrast, Figure~\ref{F:cc_power} shows the results for models T, TF, MT, and MTF. We see that the initial large-scale variation in metal density is redistributed to many different scales by the turbulence, and the resulting distribution becomes in fact white noise.  Furthermore, this process does not depend on the injection wavelength, at least in the range $L/8 \le \lambda_\mathrm{inj} \le L$ simulated in our models.

The mixing of metals driven by thermal instability is equally effective among all of our thermally unstable disks.  By comparing Figure~\ref{F:cc1_mag} with Figure~\ref{F:cc1_nomag} (or Figure~\ref{F:cc2_mag} with Figure~\ref{F:cc2_nomag}), the metal tracer fields do not exhibit noticeable differences between models~MT and~T, indicating that magnetic fields play little role in limiting the redistribution of metals by the turbulence.  As shown in the same figures, although the mixing of metals is less effective in the pre-shock region when spiral forcing is present, the mixing process is significantly accelerated near the shock front, resulting again in white noise in the aftershock region (Figure~\ref{F:cc_power}).  Therefore, we determine the turbulence induced by thermal instability in our models is the only major mechanism in driving mixing of metals, and this mechanism can effectively redistribute metals into white noise within less than inter-arm distances.

\section{QUANTIFYING THE MIXING PROCESS} \label{S:mixing}

\subsection{Diffusion Coefficients} \label{SS:dc}

Having established that the turbulence driven by thermal instability is the primary mechanism for mixing the metals, irrespective of the existence of spiral forcing and/or magnetic fields, we focus our attention on our model~T and attempt to quantify the mixing process.  It is not clear yet if this turbulent mixing can be described as a diffusion process and, if so, what diffusion coefficient describes it.  In principle, these questions could be investigated by, for instance, the recently developed test-field method \citep{BSV09,MB10}.  However, we defer this more comprehensive analysis and present a toy model for an order-of-magnitude estimate of the mixing strength and timescale.

\begin{figure}[!tbp]
\begin{center}
\plotone{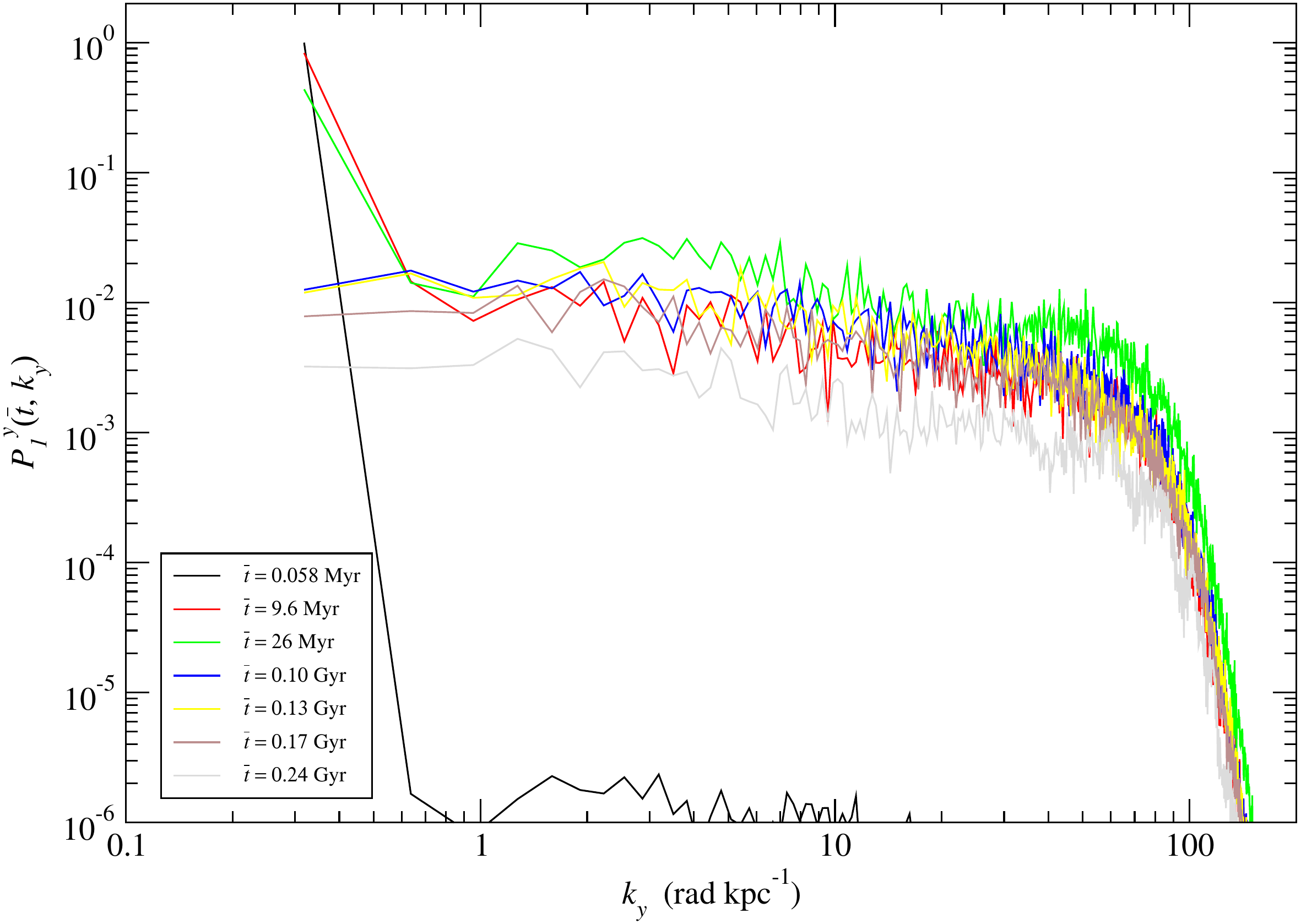}
\caption{Power spectrum in the $y$ direction at different advection times $\bar{t}$, of the metal tracer field with an injection wavelength of $\lambda_\mathrm{inj} = L$ from model~T.  Each spectrum is averaged over 10 snapshots at regular (physical) time interval from $t = 6P$ to $15P$.}
\label{F:py_cc1}
\end{center}
\end{figure}

We start by considering an observer who co-moves with the background advection (Equation~\eqref{E:cirvel}) and measures the distribution of the metals in the $y$ direction, after the turbulent flow has reached its statistically steady state.  We define $\bar{t} \equiv \left(x - x_0\right) / v_{c,x}$ as the advection time, where $x_0 = -L/2$ is the $x$ coordinate of the left boundary, and at every position $x$ and thus advection time $\bar{t}$ we compute the one-dimensional Fourier transform $\tilde{\Sigma}_X^y$ of $\Sigma_X$ in the $y$ direction. From this we compute the one-dimensional power spectrum
\begin{equation} \label{E:py}
  P_X^y(\bar{t}, k_y) = |\tilde{\Sigma}_X^y(\bar{t}, k_y)|^2.
\end{equation}
We plot the result at several values of $\bar{t}$ for $\Sigma_1$ in Figure~\ref{F:py_cc1}.  At small $\bar{t}$ the metal distribution is very close to the sinusoidal one injected from the left boundary, and thus the power spectrum shows that almost all the power is in a single, long-wavelength mode.  As the observer moves to the right, turbulent mixing redistributes the metals into many different scales while attenuating the amplitude of the initial distribution in the process.  In time, the distribution becomes white noise and the power at all scales decays roughly synchronously while the gas flows towards the right boundary.

If the process of redistributing metals were truly a diffusion process
in the $y$ direction of the observer's frame, then the distribution of metal tracers as a function of $\bar{t}$ and $y$ would obey
\begin{equation} \label{E:diffus}
  \frac{\partial\Sigma_X}{\partial\bar{t}} =
  \frac{\partial}{\partial y}\left(D\frac{\partial\Sigma_X}{\partial y}\right),
\end{equation}
where $D$ is the diffusion coefficient.\footnote{Note that we have not shown that turbulent mixing of metals really is a diffusion process, and indeed it is probably more complex than that. However, parameterizing in terms of a diffusion coefficient still provides a useful guide to the strength of the effect.}  If we assume $D$ is a constant, the solution to Equation~\eqref{E:diffus} with an initial sinusoidal distribution of wavenumber $k_\mathrm{inj} = 2\pi / \lambda_\mathrm{inj}$ is
\begin{equation}
  \Sigma_X(\bar{t},y) = \psi(\bar{t})\sin\left(k_\mathrm{inj}y\right),
\end{equation}
where
\begin{equation}
  \psi(\bar{t}) = \psi_0\exp\left(-\bar{t} / \tau_D\right),
\end{equation}
in which $\psi_0 \equiv 1$ is the amplitude of the injected distribution from the left boundary and $\tau_D = 1 / D k_\mathrm{inj}^2$ is the time constant.  Therefore, the power of the metal distribution at the injected wavelength decays exponentially according to $P^y_X(\bar{t},k_\mathrm{inj}) = \psi^2(\bar{t}) = \psi_0^2\exp\left(-2\bar{t}/\tau_D\right)$.

\begin{figure}[!tbp]
\begin{center}
\plotone{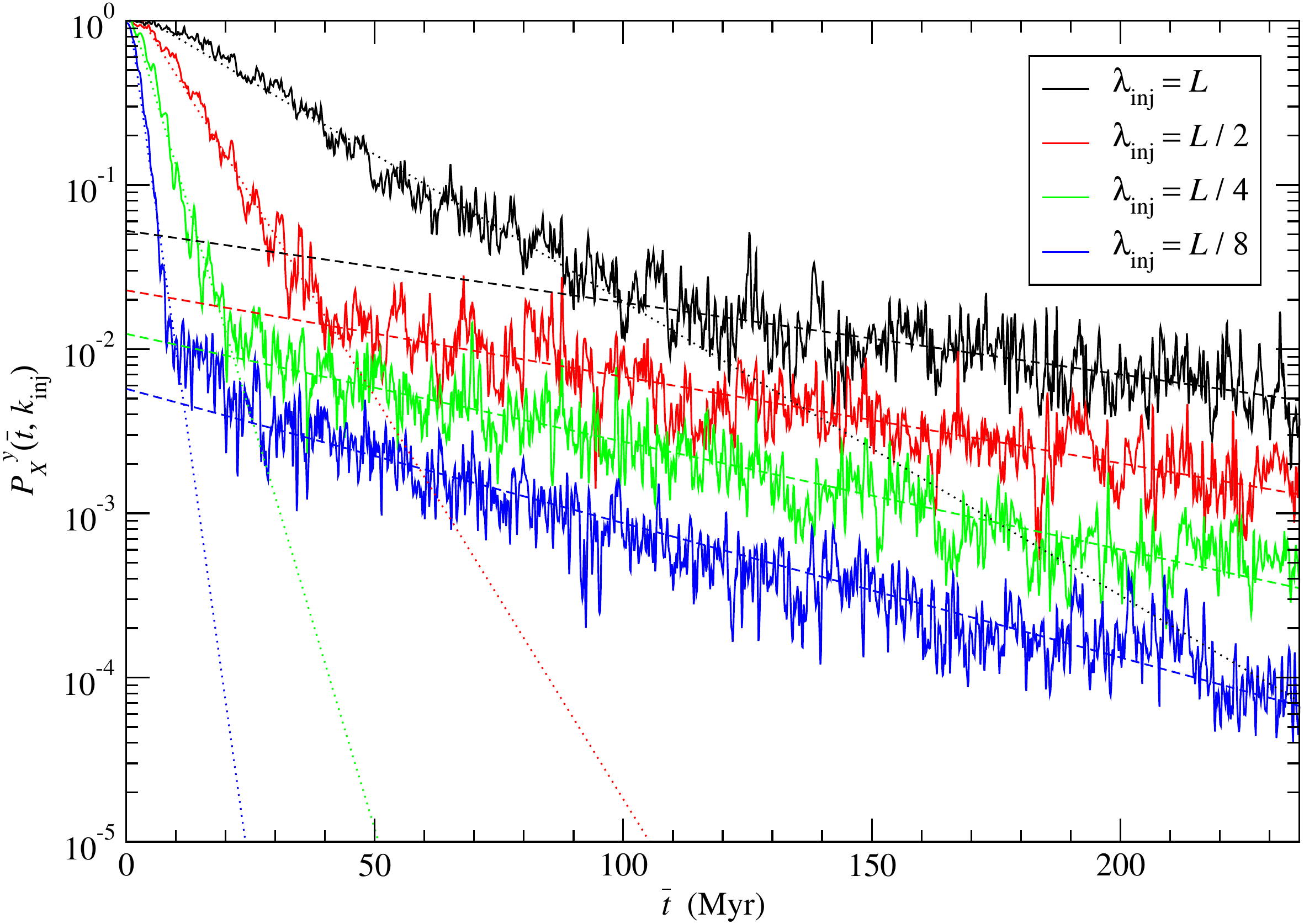}
\caption{Power of the metal tracer field at the injection wavelength $\lambda_\mathrm{inj}$ as a function of the advection time $\bar{t}$ in model~T.  Four tracer fields with different $\lambda_\mathrm{inj}$ are shown by solid lines.  The power is averaged over 10 snapshots at regular (physical) time interval from $t = 6P$ to $15P$.  The dotted lines are the regression fit of an exponential function for the first stage of the mixing process, while the dashed lines are that for the second stage of the process.}
\label{F:diffus}
\end{center}
\end{figure}

Figure~\ref{F:diffus} plots the power of the metal distribution in the $y$ direction at the injection wavenumber $P^y_X(\bar{t},k_\mathrm{inj})$ as a function of the advection time $\bar{t}$ in our model~T.  Two distinct stages of exponential decay can be seen for each metal tracer field, the first of which is steeper than the second.  The transition time $\bar{t}_0$ between the two stages marks the time required for the metal distribution to become white noise, i.e., well mixed due to the turbulence.  The shorter the wavelength of the injected distribution $\lambda_\mathrm{inj}$, the faster the metals are mixed.  With $\bar{t}_0$ identified for each tracer field, the decay of the power at each stage can be fitted separately by an exponential function as shown by the straight lines in Figure~\ref{F:diffus}, and the resulting slopes can be converted into the decay time constant $\tau_D$ and the diffusion coefficient $D$ by the formulae given above.  The numerical values of $\bar{t}_0$, $\tau_D$, and~$D$ for each $\lambda_\mathrm{inj}$ in our model~T are listed in Table~\ref{T:mixing}.

\subsection{Implications for Chemical Evolution of Disk Galaxies}

The timescales we have measured in our model~T indicate that turbulent mixing of metals driven by thermal instability is an efficient process, especially for the first stage discussed above.  The time required to eradicate kpc-scale variations in metals, i.e., $\bar{t}_0$ in Table~\ref{T:mixing}, is short compared to the orbital timescale $P$.  As we have mentioned in Section~\ref{SSS:metals}, the sinusoidal distribution of metal tracers we inject along the left boundary can be considered as the metal enrichment powered by supernovae along a spiral arm.  If the characteristic separation between star forming sites along a spiral arm is on the order of one kpc, the metals they produce will become well mixed after  $\sim$30~Myr of advection.

The second stage of turbulent mixing we see in our models is probably more relevant to long-term chemical evolution of disk galaxies.  At this stage, metals are randomly distributed in the ISM and are constantly transported by large-scale convective motions of the gas.  Like what occurs at the first stage, the signals of the metal variations at all wavelengths decay exponentially with time, although somewhat more slowly.  The decay time constant $\tau_D$ is on the order of $\sim$100~Myr and is relatively insensitive to wavelength.  Therefore, if there exists any metallicity gradient on a kilo-parsec scale, the gradient should be $e$-folded in roughly the same timescale, and this timescale is comparable to but still less than the orbital timescale of the galaxy.

In this regard, turbulent mixing of metals should be an important physical process in chemical evolution of disk galaxies, and should be included in chemical evolution models.  Although the toy model for turbulent mixing we presented in the previous section may not quantitatively describe the full dynamics of metal transport in turbulent ISM, we should have captured an order-of-magnitude estimate of the mixing strength.  The diffusion operation along with the diffusion coefficient we have measured may serve as a simple starting point for a sub-grid recipe in chemical evolution models and cosmological simulations.

We note that the diffusion coefficient of the second stage we find for kpc-scale distributions is on the same order of $c_{s,0}H \simeq 0.7$~kpc$^2$~Gyr$^{-1}$, even though the gas disk is not self-gravitating and presumably has a low \citet{SS73} $\alpha$ parameter.  In fact, $\alpha \equiv \langle\Sigma u_x u_y\rangle / \Sigma_0 c_{s,0}^2$ in our model~T is about $10^{-2}$, where $\langle \rangle$ denotes the spatial average of the quantity enclosed.  This demonstrates that the transport of metals does not strictly follow the viscous evolution of the gas disk.  The convective motion of the gas can actually carry the metals over larger distances than a pure viscous stress allows.  Therefore, the assumption of the same $\alpha$ prescription for both the gas and the metals in a chemical evolution model is not correct.

\section{CONCLUSIONS}

In this work, we simulate a local patch of a vertically thin disk galaxy and study the transport of metals  with a variety of physical conditions.  Specifically, we investigate the ability of thermal instability, spiral shocks, and/or magnetic fields to homogenize metals.  We find that turbulence driven by thermal instability is especially effective in mixing the metals, regardless of the presence or absence of spiral shocks and magnetic fields.

We observe two different modes of turbulent mixing in our thermally unstable disks.  The first mode is for the turbulent gas to stir large-scale variations of metals into a random distribution.  The timescale for this mode is short compared to the local orbital time in the galaxy, and this mode may contribute to obliterate the chemical inhomogeneities introduced by star forming activities along spiral arms.  The second mode is for randomly-distributed metals to be continually homogenized over time by the turbulence.  We find the timescale for this process is relatively insensitive to wavelength and is on the order of half the orbital timescale.  This mode of turbulent mixing, therefore, should be of  significance in reducing the metallicity gradient in a disk galaxy.

We find that turbulent mixing of metals driven by thermal instability is more efficient than what a simple \citet{SS73} $\alpha$ prescription of viscosity for the gas would suggest.  The convective motion of the turbulent gas can in fact transport metals over larger distances, especially for kpc-scale variations.  The dynamics is perhaps more complicated than ordinary diffusive transport with a constant coefficient.  In an attempt to capture its qualitative behavior, however, we have devised a toy prescription in terms of a wavelength-dependent diffusion coefficient and measured its numerical values for our model galactic disk.  In principle, this prescription could be adopted as a sub-grid physical process in semi-analytic chemical evolution models as well as cosmological simulations.  Doing so should help us further constrain the dynamical history of disk galaxies.

\acknowledgments
This work was supported by the Alfred P.\ Sloan Foundation, the NSF through grant CAREER- 0955300, and NASA through Astrophysics Theory and Fundamental Physics Grant NNX09AK31G, and a Chandra Space Telescope Grant. The simulations presented in this paper were conducted using the supercomputing system Pleiades at the University of California, Santa Cruz.

\appendix
\section{HYPERDIFFUSION AND SHOCK DIFFUSION WITH FIXED MESH REYNOLDS NUMBER} \label{S:fixRe}

Depending on the system of interest, the Pencil Code requires artificial terms in all dynamical equations, except those describing the passive scalar fields, to stabilize the scheme.  To simulate transonic turbulence with formation of shocks, we include hyper-diffusion and shock diffusion terms in our simulations. The hyper-diffusion terms use sixth-order derivatives to damp numerical noise at high wavenumber but preserve power on larger scales \citep{HB04,JK05}, while the shock diffusion terms are of von Neumann type \citep{HBM04,LJ08}.  The usual approach is to set the diffusion coefficients to constant values ($\nu_3$ and $a_s$ defined below).  However, we have implemented a new strategy to dynamically adjust them so that the mesh Reynolds number remains nearly constant.  We briefly describe the underlying concept of this implementation in this section.

The hyper-diffusion terms are of the form
\begin{equation} \label{E:hypdiff}
  \nu_3\left(\frac{\partial^6\mathcal{Q}}{\partial x^6}
           + \frac{\partial^6\mathcal{Q}}{\partial y^6}
           + \frac{\partial^6\mathcal{Q}}{\partial z^6}\right),
\end{equation}
where $\mathcal{Q}$ is the primitive variable to be solved for and $\nu_3$ is the hyper-diffusion coefficient.  The strength of this operation can in fact be evaluated by comparing Equation~\eqref{E:hypdiff} with the advection term $\vec{u}\cdot\del\mathcal{Q}$.  Consider a specific signal (or rather, noise) in $\mathcal{Q}$ at wavenumber $\vec{k}$.  It is damped faster than being advected away if
\begin{equation} \label{E:hypcrit}
  |\vec{u}\cdot\vec{k}| \lesssim \nu_3 (k_x^6 + k_y^6 + k_z^6).
\end{equation}
Since $|\vec{u}\cdot\vec{k}| \leq u k \leq u_\mathrm{max} k$ and $k_x^6 + k_y^6 + k_z^6 \sim k^6$, where $u_\mathrm{max}$ is the maximum magnitude of velocity $\vec{u}$ in the computational domain, Equation~\eqref{E:hypcrit} implies $u_\mathrm{max} \lesssim \nu_3 k^5$.  We define the mesh Reynolds number for hyper-diffusion as
\begin{equation} \label{E:reh}
  \mathrm{Re}_h \equiv \frac{u_\mathrm{max}}{\nu_3 k_\mathrm{Nyq}^5},
\end{equation}
where $k_\mathrm{Nyq} \equiv \pi / \max(\delta x, \delta y, \delta z)$ is the Nyquist wavenumber, and $\delta x$, $\delta y$, and $\delta z$ are grid spacing in the $x$, $y$, and $z$ directions, respectively.  The aforementioned criterion for damping signals at Nyquist frequency then becomes $\mathrm{Re}_h \lesssim 1$.

Motivated by this criterion, we invert Equation~\eqref{E:reh} to find the value of a time-dependent, spatially uniform hyper-diffusion coefficient $\nu_3$ with a fixed mesh Reynolds number $\mathrm{Re}_h$:
\begin{equation} \label{E:nu3}
  \nu_3 = \nu_3(t) = \frac{u_\mathrm{max}(t)}{k_\mathrm{Nyq}^5 \mathrm{Re}_h}.
\end{equation}
In other words, we determine the maximum magnitude of the velocity field $u_\mathrm{max}$ at the beginning of each time step, and use this information to assign for this step the value of the diffusion coefficient $\nu_3$ calculated from Equation~\eqref{E:nu3}.  This way, we maintain the artificial diffusion at Nyquist frequency with roughly the same strength.  Due to the high-order dependence of the hyper-diffusion operator on wavenumber ($\sim k^6$), the damping of noise is then concentrated at and near the Nyquist frequency while quickly diminishing towards longer wavelengths.

Similarly, we can control the strength of shock diffusion by fixing the appropriately-defined corresponding mesh Reynolds number.  The shock diffusion terms are of the form $\del\cdot(\nu_s\del\mathcal{Q})$ except the one for the momentum equation, which is written as a bulk viscosity $\rho^{-1}\del\left(\rho\nu_s\del\cdot\vec{u}\right)$.  The diffusion coefficient $\nu_s$ is of the form $\nu_s = a_s \max(-\del\cdot\vec{u},0)$, where $a_s$ is a positive constant; $\nu_s$ is thus spatially variable and is proportional to the local convergence of the flow.  Consider again a signal in $\mathcal{Q}$ with wavenumber $\vec{k}$ and compare the strength of shock diffusion with that of the advection.  One obtains
\begin{equation}
  |\vec{u}\cdot\vec{k}| \la \nu_s\left(k_x^2 + k_y^2 + k_z^2\right) = \nu_s k^2.
\end{equation}
Note that this criterion is only meaningful near shock fronts.  We therefore define the mesh Reynolds number for shock diffusion as
\begin{equation} \label{E:res}
  \mathrm{Re}_s \equiv \frac{\max\left(\left(\left|u_x k_x\right|
                                     + \left|u_y k_y\right|
                                     + \left|u_z k_z\right|\right) / k^2\right)}
                            {a_s\max(-\del\cdot\vec{u})},
\end{equation}
which is the most conservative measurement of the Reynolds number at the strongest local convergence of the flow.\footnote{The Reynolds number $\mathrm{Re}_s$ is undefined if there is no position for which $\del\cdot\vec{u} < 0$, and no shock diffusion operates in this case.}  With this definition, then, we solve Equation~\eqref{E:res} for the value of the constant $a_s$ with a fixed Reynolds number $\mathrm{Re}_s$ at the beginning of each time step.

In all of our simulations, we use $\mathrm{Re}_h = \mathrm{Re}_s = 1/4$ except for model~MTF, in which we use $\mathrm{Re}_s = 1/6$.

\section{RESOLUTION STUDY} \label{S:res}

\begin{figure}[!tbp]
\begin{center}
\epsscale{.8}
\plotone{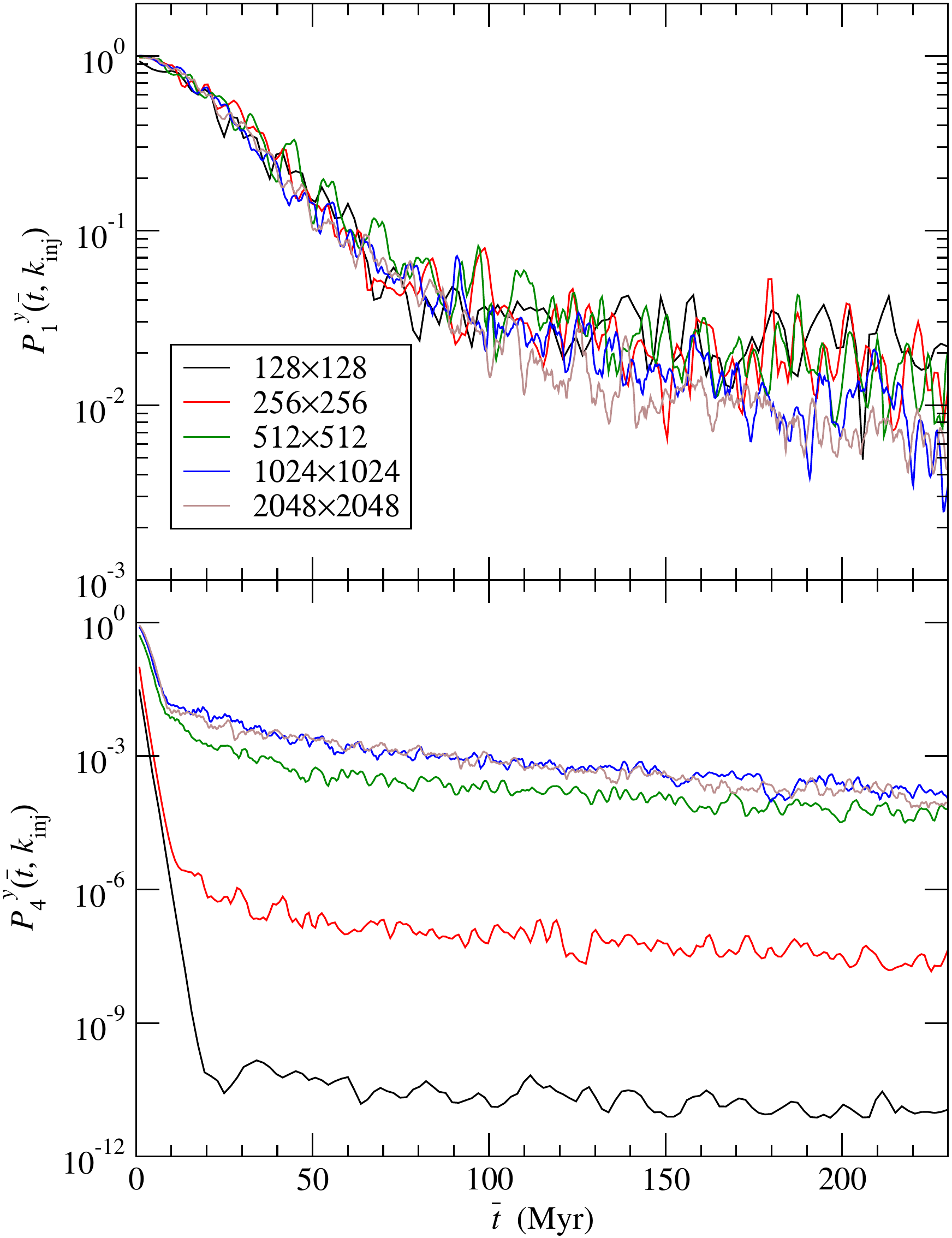}
\caption{Power of the metal tracer field at the injection wavelength $\lambda_\mathrm{inj}$ for $\lambda_\mathrm{inj} = L$ (\emph{top}) and $\lambda_\mathrm{inj} = L / 8$ (\emph{bottom}) as a function of the advection time $\bar{t}$ in model~T.  The power is averaged over 10 snapshots at regular (physical) time interval from $t = 6P$ to $15P$.  Different lines are the power from the same model at different resolutions, and the lines are smoothed by running averages to emphasize their general trend with respect to $\bar{t}$.}
\label{F:res}
\end{center}
\end{figure}

In this section, we demonstrate that our simulations are numerically converged.  The diagnostic we choose to present is the $y$-power $P_X^y(\bar{t}, k_y)$ defined in Equation~\eqref{E:py}, which is arguably the most important measurement from our simulations made in this paper.  Figure~\ref{F:res} plots the power $P_X^y(\bar{t}, k_\mathrm{inj})$ as a function of the advection time $\bar{t}$ for $\lambda_\mathrm{inj} = L$ and $\lambda_\mathrm{inj} = L / 8$ from model~T at different resolutions.  For the case of $\lambda_\mathrm{inj} = L$, the curves from resolutions 128$\times$128 to 2048$\times$2048 are roughly on top of each other, and thus they all exhibit almost the same behavior on the two stages of turbulent mixing discussed in Section~\ref{S:mixing}.  For the case of $\lambda_\mathrm{inj} = L / 8$, significant amounts of power are lost in the low resolution simulations due to their inability to resolve the short wavelength of the injected metal distribution.  However, one can see in Figure~\ref{F:res} that the difference between each pair of adjacent curves decreases with higher resolutions, and the curves for the highest two resolutions, 1024$\times$1024 and 2048$\times$2048, roughly coincide, indicating numerical convergence.  Therefore, turbulent mixing of metals in our simulations should not be dominated by numerical dissipation, and the values listed in Table~\ref{T:mixing} should be robust.

\begin{deluxetable}{lcl}
\tablecaption{Adopted Physical Parameters\label{T:param}}
\tablewidth{0pt}
\tablehead{
\colhead{Parameter} & \colhead{Symbol} & \colhead{Value}
}
\startdata
Galactocentric distance & $R_0$ & 10~kpc \\
Angular circular speed & $\Omega_0$ & 26~km~s$^{-1}$~kpc$^{-1}$ \\
Orbital period & $P$ & 240~Myr \\
Spiral-arm multiplicity & $m$ & 2 \\
Spiral-arm pitch angle & $i$ & 5.7\degr \\
Inter-arm distance & $L$ & 3.1~kpc \\
Initial Toomre stability parameter & $Q_0$ & 1.5 \\
Mean molecular weight & $\mu$ & 1 \\
Vertical disk scale height & $H$ & 95~pc \\
\enddata
\end{deluxetable}

\begin{deluxetable}{lcccccccc}
\tablecaption{List of Models\label{T:model}}
\tablewidth{0pt}
\tablehead{
\colhead{Model}
& \colhead{Forcing\tablenotemark{a}}
& \colhead{Equation of State\tablenotemark{b}}
& \colhead{Magnetized\tablenotemark{c}}
& \colhead{Highest Resolution}
}
\startdata
Control & No & Isothermal & No & 1024$\times$1024 \\
F & Yes & Isothermal & No & 1024$\times$1024 \\
T & No & Non-isothermal & No & 2048$\times$2048 \\
TF & Yes & Non-isothermal & No & 2048$\times$2048 \\
M & No & Isothermal & Yes & 1024$\times$1024 \\
MF & Yes & Isothermal & Yes & 1024$\times$1024 \\
MT & No & Non-isothermal & Yes & 2048$\times$2048 \\
MTF & Yes & Non-isothermal & Yes & 1024$\times$1024 \\
\enddata
\tablenotetext{a}{If forcing exists, $F = 3$\%; $F = 0$, otherwise.}
\tablenotetext{b}{For isothermal disks, $\Sigma_0 = 13~M_\sun$~pc$^{-2}$ and $c_{s,0} = 7.0$~km~s$^{-1}$.  For non-isothermal disks, $\Sigma_0 = 12~M_\sun$~pc$^{-2}$, $c_{s,0} = 6.4$~km~s$^{-1}$, and $\gamma = 1.8$.}
\tablenotetext{c}{For magnetized disks, $\beta_0 = 2$.}
\end{deluxetable}
\begin{deluxetable}{ccccccc}
\tablecaption{Properties of the Mixing Process for Different Metal Tracers in Model~T\label{T:mixing}}
\tablewidth{0pt}
\tablehead{
\multicolumn{2}{c}{} &
\multicolumn{2}{c}{First Stage} & &
\multicolumn{2}{c}{Second Stage}\\ \cline{3-4} \cline{6-7}
\colhead{$\lambda_\mathrm{inj}$} & \colhead{$\bar{t}_0$} &
\colhead{$\tau_D$} & \colhead{$D$}                  & &
\colhead{$\tau_D$} & \colhead{$D$}\\
\colhead{(kpc)}                  & \colhead{(Myr)}       &
\colhead{(Myr)}    & \colhead{(kpc$^2$~Gyr$^{-1}$)} & &
\colhead{(Gyr)}    & \colhead{(kpc$^2$~Gyr$^{-1}$)}
}
\startdata
3.1\phn &    100 &    48\phd\phn & 5.2\phn & & 0.20 & 1.2\phn\phn\\
1.6\phn & \phn41 &    18\phd\phn & 3.5\phn & & 0.16 & 0.38\phn\\
0.78    & \phn22 & \phn8.6       & 1.8\phn & & 0.13 & 0.12\phn\\
0.39    & \phn12 & \phn4.0       & 0.96    & & 0.11 & 0.037
\enddata
\tablecomments{$\lambda_\mathrm{inj}$ is the wavelength of the metal distribution injected from the left boundary.  $\bar{t}_0$ denotes the approximate advection time when the mixing process transitions from the first stage to the second.  $\tau_D$ and $D$ respectively represent the decay time constant of the injected distribution and the corresponding diffusion coefficient at each stage.}
\end{deluxetable}
\end{document}